\documentstyle[graphicx]{mn2e}
\footnotesize
\newdimen\digitwidth    
\setbox0=\hbox{\rm0}
\digitwidth=\wd0
\catcode`!=\active
\def!{\kern\digitwidth}
\normalsize
\title[TEMPO2, a new pulsar timing package. I: Overview] 
{TEMPO2, a new pulsar timing package. I: Overview}
\author[G. Hobbs et al.]
{G. B. Hobbs,\thanks{Email: george.hobbs@csiro.au}
R. T. Edwards,
R. N. Manchester
\\
Australia Telescope National Facility, CSIRO, PO~Box~76, Epping,
NSW~1710, Australia \\}

\date{}
\begin{document}
\maketitle
\newcommand{\setthebls}{
}
\setthebls
\begin{abstract}

Contemporary pulsar timing experiments have reached a sensitivity
level where systematic errors introduced by existing analysis
procedures are limiting the achievable science. We have developed
\textsc{tempo2}, a new pulsar timing package that contains propagation
and other relevant effects implemented at the 1\,ns level of precision
(a factor of $\sim 100$ more precise than previously obtainable). In
contrast with earlier timing packages, \textsc{tempo2} is compliant
with the general relativistic framework of the IAU 1991 and 2000
resolutions and hence uses the International Celestial Reference
System, Barycentric Coordinate Time and up-to-date precession,
nutation and polar motion models.  \textsc{Tempo2} provides a generic
and extensible set of tools to aid in the analysis and visualisation
of pulsar timing data. We provide an overview of the timing model, its
accuracy and differences relative to earlier work. We also present a
new scheme for predictive use of the timing model that removes existing
processing artifacts by properly modelling the frequency dependence of
pulse phase.

\end{abstract}

\section{Introduction}

 Pulsar timing observations have produced some of the most exciting
 results in pulsar astronomy and indeed in all of astronomy. For
 instance, such results have included the first detection of
 extra-solar planets (Wolszczan \& Frail, 1992)\nocite{wf92},
 stringent tests of the general theory of relativity (e.g. Stairs
 2003\nocite{sta03}), revealed dispersion measure variations due to
 the interstellar medium (e.g. Backer et al. 1993)\nocite{bhh93},
 pulsar proper motions (e.g. Hobbs et al. 2004)\nocite{hlk+04} and
 irregularities in the spin-down of pulsars (e.g. Lyne
 1999\nocite{lyn98b}). Pulsar timing is now being used to verify
 terrestrial time standards and the Solar System ephemeris and in
 searches for gravitational radiation (see for example, Foster \&
 Backer 1990\nocite{fb90} and Jenet et al. 2005\nocite{jhlm05}).

 An overview of pulsar timing has been given by numerous authors (for
 example, Manchester \& Taylor 1977, Backer \& Hellings 1986, Lyne \&
 Smith 1998 and Lorimer \& Kramer
 2005)\nocite{mt77}\nocite{bh86}\nocite{ls98}\nocite{lk05}.  In brief,
 the arrival times of pulses (TOAs) are measured at a radio
 observatory for a particular pulsar over many years.  These TOAs need
 adjustment so that they represent arrival times in an inertial
 reference frame.  This is accomplished by transforming each measured
 arrival time to an arrival time in the reference frame of the pulsar
 by first calculating arrival times at the Solar System barycentre
 (SSB) and then, if necessary, including additional terms required to
 model the pulsar's orbital motion. A model of the pulsar's spin-down
 behaviour, the ``timing model'' or ``timing ephemeris'', is fitted to
 these arrival times.  If significant systematic deviations are seen
 when calculating the differences between the actual arrival times and
 the best-fit model arrival times (known as timing residuals) then it
 is clear that the model is not fully describing the true pulsar
 parameters; a positive residual corresponds to the pulse arriving
 later than predicted. Such discrepancies can be due to many effects
 including unmodelled binary companions or binary parameters,
 irregularities in the spin-down of the pulsar, or poor estimation of
 the astrometric or rotational parameters. For instance, an incorrect
 estimate of the pulsar's position or its proper motion leads to a
 poor determination of the barycentric arrival times, which will
 produce a sinusoidal feature in the timing residuals.  This timing
 technique therefore allows pulsar parameters to be measured extremely
 precisely; the precision improves with longer data sets and more
 accurate TOA measurements.

 Both the conversion from the measured TOAs to barycentric arrival
 times and the model fitting required to obtain precise pulsar
 parameters are complex and can only be carried out within a computer
 program.  Programs such as \textsc{psrtime} at Jodrell Bank
 observatory, \textsc{timapr} at Bonn, \textsc{antiope} at Nancay and
 \textsc{cphas} at Hartebeesthoek observatories have already been
 developed.  However, the most widely used and best known package is
 \textsc{tempo} which has been maintained and distributed by Princeton
 University and the Australia Telescope National
 Facility\footnote{http://www.atnf.csiro.au/research/pulsar/tempo}.
 This package is extremely powerful, but the algorithms implemented
 are poorly documented and only provide a timing precision of
 $\sim$100\,ns. Recent high precision timing experiments produce
 root-mean-square (rms) residuals of this order and therefore such
 results are systematically affected by inaccuracies in the
 \textsc{tempo} algorithms. A further limitation of \textsc{tempo} is
 that it can only be used to analyse one pulsar at a time. In order to
 study the recently discovered double-pulsar system \cite{lbk+04}, to
 search for gravitational waves or to look for irregularities in
 terrestrial time standards, it is advantageous to analyse multiple
 pulsars simultaneously. We have developed a new package, known as
 \textsc{tempo2}, which is based on the original \textsc{tempo}
 (hereafter called \textsc{tempo1}), but has a significant number of
 new and improved features.

%


 The aim of this paper is not to provide a user manual, but rather to
 1) give a succinct description of the algorithms implemented, 2)
 highlight features that are not available in existing timing packages
 and 3) describe the accuracy of \textsc{tempo2}. Full documentation
 and download instructions for \textsc{tempo2} can be obtained from
 our web
 site\footnote{http://www.atnf.csiro.au/research/pulsar/tempo2}.
 Details of the algorithms used in \textsc{tempo2} in order to achieve
 accuracies of 1\,ns will be presented in Paper~II of this series.
 Methods to simulate the effects of gravitational waves on pulsar
 timing data and utilities to place limits on the existence of a
 gravitational wave background will be described in Paper~III.

 In \S 2 we describe real and simulated pulse arrival times used for
 testing and demonstrating the various features of \textsc{tempo2}.
 \S 3 provides a description of the conversion between site arrival
 times to arrival times in the pulsar frame through the use of clock
 correction files, propagation delays and a planetary ephemeris.  The
 fitting algorithms implemented in \textsc{tempo2} for single datasets
 are described in \S 4.  \S 5 describes analysis methods for the
 fitted parameters and their uncertainties and \S 6 contains
 information on \textsc{tempo2} routines to study the resulting timing
 residuals.  \textsc{Tempo2} provides a predictive facility which is
 described in \S 7.

\section{Real and simulated pulse arrival times}

 The \textsc{tempo2} software is based around 1) an ``engine'' that
 calculates the barycentric arrival times, forms the timing residuals
 and carries out the weighted least-squares fit and 2) ``plug-ins''
 that add to the functionality of \textsc{tempo2} and allow the
 results to be analysed and presented in a user-friendly form.  For
 instance, a plug-in is available to plot the timing residuals of
 multiple pulsars simultaneously, another to determine the power
 spectrum of the residuals and another to graph the clock corrections
 that \textsc{tempo2} is applying to the measured arrival times. A
 full listing of the currently available plug-ins is provided in the
 Appendix.

 It is now common to combine TOAs obtained at different observatories
 with different back-end systems and receivers.  These almost
 invariably give rise to a constant offset or ``jump'', between each
 set of TOAs.  \textsc{Tempo2} can fit for such jumps between
 observations at different telescopes, with different observing
 frequencies or back-end systems, between a range of dates, or on any
 other given parameter.  This is made possible by a new free format
 for the measured pulse arrival times.  This free format allows
 additional flags providing user-definable parameters such as the
 backend system, the observation length or the observation bandwidth.
 Using the graphical plug-in features it is also possible to plot the
 pre- and post- fit timing residuals versus time or other parameters
 such as the observation length, parallactic angle or attenuation
 settings.

\begin{figure}
  \includegraphics[width=6cm,angle=-90]{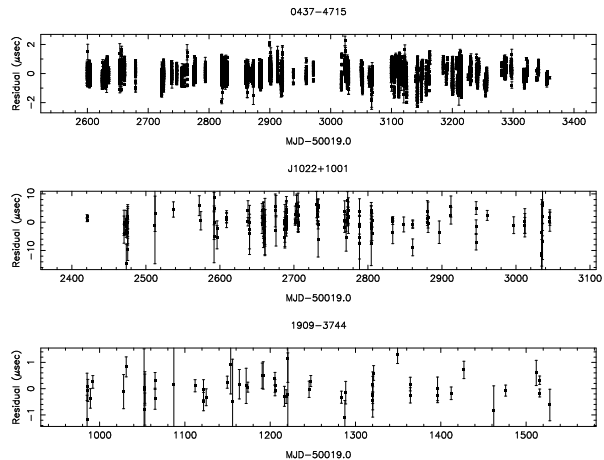}
  \caption{The timing residuals for PSRs~J0437$-$4715, J1022$+$1001
 and J1909$-$3744 observed as part of the Parkes Pulsar Timing Array
 project. This figure was produced using the \textsc{splk} interface
 to \textsc{tempo2}.}\label{fg:selpsrs}
 \end{figure}

 It is essential to test the algorithms implemented within
 \textsc{tempo2} with precise TOAs.  We have selected three pulsars,
 listed in Table~\ref{tb:psrs} that have been observed for the Parkes
 Pulsar Timing Array project
 (PPTA)\footnote{http://www.atnf.csiro.au/research/pulsar/ppta}
 which aims to detect gravitational waves by looking for correlated
 signatures in the timing residuals of multiple pulsars (Jenet et
 al. 2005)\nocite{jhlm05}.  Timing residuals for these pulsars are
 shown in Figure~\ref{fg:selpsrs}.  PSRs~J0437$-$4715 and J1909$-$3744
 provide high quality timing observations at the sub-500\,ns
 level. PSR~J1022$+$1001 has an ecliptic latitude of $-$0.06$^\circ$
 and hence, the TOAs are affected by Solar System dispersion and
 Shapiro delays. Full details of the project and observing details
 will be described in a later paper.

 \begin{table}\setlength{\tabcolsep}{4.5pt}
  \caption{A selection of the pulsars observed in the Parkes Timing
  Array Project.  We list the pulsars' names, pulse frequencies
  ($\nu$), observing spans, numbers of TOAs (N$_{\rm TOA}$), observing
  frequencies ($f$) and the post-fit rms residuals
  (rms).}\label{tb:psrs}
  \begin{center}
  \begin{tabular}{llllll} \hline
   \multicolumn{1}{c}{PSR} & \multicolumn{1}{c}{$\nu$} & \multicolumn{1}{c}{Span} &\multicolumn{1}{c}{N$_{\rm TOA}$} & \multicolumn{1}{c}{f}   & \multicolumn{1}{c}{rms}      \\ 
       & \multicolumn{1}{c}{(Hz)}  & \multicolumn{1}{c}{(d)}  &           & \multicolumn{1}{c}{(MHz)}& \multicolumn{1}{c}{($\mu$s)} \\ \hline
   J0437$-$4715 & 173.6879 & 761 & 6382 & 1340 & 0.49  \\
   J1022$+$1001 &  60.7794 & 600 & 142 & 3100/1400/685 & 3.7 \\
   J1909$-$3744 & 339.3157 & 542 & 56 & 3100 & 0.35 \\
  \hline\end{tabular}
  \end{center}
 \end{table}

\begin{figure}
  \includegraphics[width=6cm,angle=-90]{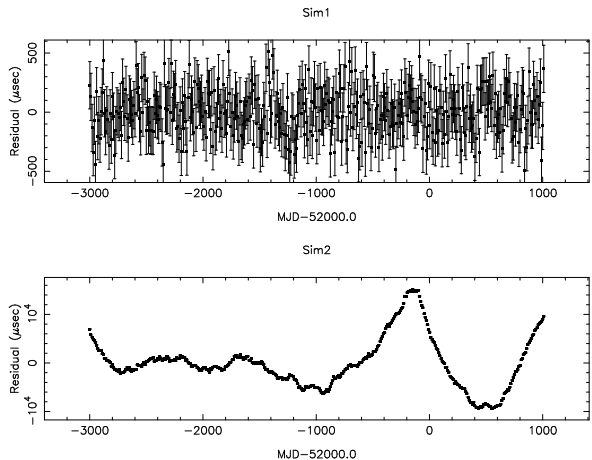}
  \caption{\textsc{Tempo2} simulation of ``white'' (upper panel) and ``red''
  (lower panel) timing residuals.  The lower panel contains both red
  timing noise simulated using a steep power-law spectrum and a small
  glitch event at MJD 51900. This figure was produced using the
  \textsc{fake} and \textsc{splk} interfaces to \textsc{tempo2}.}\label{fg:simulate}
 \end{figure}

 For more detailed tests we use a \textsc{tempo2} plug-in capable of
 simulating pulsar timing residuals in the presence of red noise or
 with glitch events (see Figure~\ref{fg:simulate}).  These TOAs are
 determined by repeatedly forming the pulsar timing residuals and then
 subtracting these residuals from the TOAs until the TOAs exactly
 match the timing model provided.  The simulated residuals are then
 output after the addition of ``white'' (Gaussian) and/or ``red''
 noise (modelled by summing many sinusoids with random phase, but with
 amplitudes given by the requested power-law spectrum).

\section{Forming the pulse emission time}

 The timing procedure starts by converting the measured topocentric
 TOAs to the pulse emission time in the pulsar frame ignoring the
 frequency independent propagation delay from the pulsar to the SSB.
 Full details of this transformation, its accuracy and differences
 relative to \textsc{tempo1} will be described in Paper~II. Here we
 summarise the transformation as
 \begin{equation} 
  \Delta t = \Delta_{\rm C} + \Delta_{\rm A} + \Delta_{\rm E_\odot} +
  \Delta_{\rm R_\odot} + \Delta_{\rm S_\odot}  - D/f^2 + \Delta_{\rm
  VP} + \Delta_{\rm B}
 \end{equation}
 where $\Delta_{\rm C}$ contains various clock corrections (see
 \S\ref{sec:clock}), $\Delta_{\rm A}$ the atmospheric propagation
 delays (\S\ref{sec:atmos}), $\Delta_{\rm E_\odot}$ the Solar System
 Einstein delay (\S\ref{sec:einstein}), $\Delta_{\rm R_\odot}$ the
 Solar System Roemer delay (\S\ref{sec:prop}), $\Delta_{\rm S_\odot}$
 the Solar System Shapiro delay (\S\ref{sec:shapiro}), $D/f^2$ models
 the dispersive component of the light travel time (\S\ref{sec:dm}),
 $\Delta_{\rm vp}$ describes the excess vacuum propagation delay due
 to secular motion (\S\ref{sec:vacuum}) and $\Delta_{\rm B}$ contains
 terms that describe any orbital motion (\S\ref{sec:binary}).  In
 Table~\ref{tb:corr} we list various effects that must be taken into
 account when forming barycentric arrival times from the observed
 TOAs.  The table also provides a typical value or range for the
 magnitude of each effect and whether or not it is included in
 \textsc{tempo1}.

 \begin{table*}
  \caption{Corrections and their typical sizes for phenomena included
  in \textsc{tempo2}.}\label{tb:corr}
  \begin{tabular}{llll} \hline
   Correction & Typical value/range & \textsc{tempo1}\\
   \hline
   Observatory clock to TT                & 1\,$\mu$s  & Y\\
   Hydrostatic tropospheric delay         & 10\,ns   & N\\
   Zenith wet delay                       & 1.5\,ns  & N\\
   IAU precession/nutation                & $\sim$5\,ns & N$^a$ \\
   Polar motion                           & 60\,ns   & N\\
   $\Delta$UT1                            & 1$\mu$s   & Y\\
   Einstein delay                         & 1.6\,ms    & Y \\
   Roemer delay                           & 500\,s  & Y \\
   Shapiro delay due to Sun               & 112\,$\mu$s & Y \\
   Shapiro delay due to Venus             & 0.5\,ns  & N\\
   Shapiro delay due to Jupiter           & 180\,ns  & N\\
   Shapiro delay due to Saturn            & 58\,ns   & N\\
   Shapiro delay due to Uranus            & 10\,ns   & N\\
   Shapiro delay due to Neptune           & 12\,ns   & N\\
   Second order Solar Shapiro delay       & 9\,ns    & N\\
   Interplanetary medium dispersion delay & 100\,ns$^b$ & Y\\
   Interstellar medium dispersion delay   & $\sim$1\,s$^b$ & Y\\
  \hline\end{tabular}

  $^a$ earlier precession/nutation model implemented \\
  $^b$ observing frequency and pulsar dependent, typical value
  for 1400\,MHz listed.
 \end{table*}

 \subsection{Clock corrections}\label{sec:clock}

  The TOAs provided to \textsc{tempo2} are recorded against local
  observatory clocks. Such clocks are typically derived from a
  precision frequency standard with good short-term stability, such as
  a hydrogen maser. On longer time scales (months to years) these
  clocks deviate significantly from uniformity and are therefore
  unsuitable for precision pulsar timing. However, it is generally
  possible to remove these errors down to the precision provided by
  the best available terrestrial time scale through the application of
  corrections derived from monitoring the offsets between pairs of
  clocks. For example, the PPTA pulsars are observed at the Parkes
  observatory where the offset between the observatory 1
  pulse-per-second signal (derived from a hydrogen maser) is compared
  both to the clock signal broadcast by Global Positioning System
  (GPS) satellites and by common-view GPS monitoring to the Australian
  national time scale UTC(AUS), maintained by the National Measurement
  Institute.  The Bureau International des Poids et Mesures (BIPM) in
  turn publishes a monthly bulletin (Circular T) tabulating offsets
  between various clock pairs. Using Circular T, measurements can be
  referred from an intermediate clock (e.g. UTC(AUS) or GPS time) to
  Universal Coordinated Time (UTC). UTC is a timescale formed through
  the weighting of data from an ensemble of atomic clocks from around
  the world.  This in turn is related to Temps Atomique International
  (TAI) by an integer number of ``leap'' seconds, which are inserted
  to maintain approximate synchrony between UTC and the irregular
  rotation of the Earth (these are announced in Bulletin C of the
  International Earth Rotation Service). TAI is the most stable
  long-term time scale available in near real-time.

  The ultimate aim of the clock correction process is to transform
  measurements into the Geocentric Celestial Reference System
  (GCRS), for which the coordinate time is denoted TCG, expressed in
  units of the SI second. Owing to their gravitational and rotational
  energy, terrestrial atomic clocks made to approximate the SI second
  do not run at the same rate as TCG. Instead, these clocks are used
  to define realizations of a timescale known as Terrestrial Time
  (TT), which differs from TCG by a constant rate in such a way that
  its unit corresponds to the SI second on the surface of the geoid.
  One possible realization of TT is obtained directly from TAI:

  \begin{equation}
   \mathrm{TT(TAI)} = \mathrm{TAI} + 32.184\;\mathrm{s},
  \end{equation}

  \noindent however TAI has instabilities and inaccuracies for which
  corrections frequently become available at a later date. The best
  available stability is currently provided by the retroactive
  timescales published by the BIPM (Guinot 1988\nocite{gui88}, Petit
  2003\nocite{pet03a}), the most recent of which is denoted
  TT(BIPM04).

  The \textsc{tempo2} framework for handling clock corrections was
  designed with maximum flexibility in mind, with the possibility of
  processing data sets with a heterogeneous collection of different
  observatories, clocks, and clock correction paths. The scheme is
  based around a database of ASCII files tabulating the offsets
  between named pairs of clocks. Given the name of the clock against
  which a TOA is measured, and the name of the realisation of TT to
  which it should be transformed, corrections can be applied based
  upon a manually or automatically determined sequence derived from
  linear interpolation of values from files found in the database.
  Step changes such as leap seconds are also possible. An ancillary
  suite of programs allows for the production of \textsc{tempo2}
  format files from external data sources such as Circular T, and
  provides capabilities for averaging, resampling and various analytic
  procedures for assessing the quality of data present.

%


\subsection{Atmospheric propagation delays}\label{sec:atmos}

\begin{figure}
   \begin{center}
    \includegraphics[angle=-90,width=7cm]{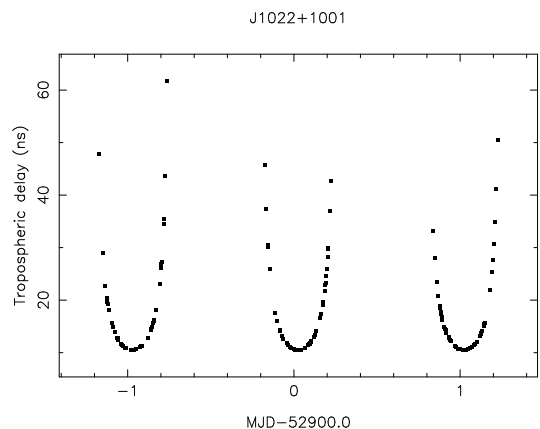}
   \end{center}
  \caption{The computed hydrostatic tropospheric delay for simulated pulse times 
of arrival, assuming a constant surface atmospheric pressure.}\label{fg:hydro}
 \end{figure}

  The group velocity of radio waves in the atmosphere differs from the
  vacuum speed of light. Refractivity is induced both by the ionised
  fraction of the atmosphere (mainly in the ionosphere) and the
  neutral fraction (mainly in the troposphere).  The tropospheric
  propagation delay can be separated into the so-called
  ``hydrostatic'' and ``wet'' components (see Paper~II).  For the
  highest timing precision, it is possible to provide \textsc{tempo2}
  with a tabulated list of surface atmospheric pressure measured at an
  observatory for the calculation of the hydrostatic delay which will
  be of the order of 10\,ns. If atmospheric pressure data are
  unavailable then \textsc{tempo2} can, if required, use a canonical
  value of one standard atmosphere. This assumption results in errors
  of the order of 1.5\,ns.  In Figure~\ref{fg:hydro} we show computed
  hydrostatic tropospheric delays for simulated TOAs for
  PSR~J1022$+$1001, assuming a constant surface atmospheric pressure
  and a $\pm 5$-h hour-angle range. Diurnal variations arise
  due to the dependence of atmospheric path length on source elevation
  (in the simulated observations the elevation varies from 6 to 46
  degrees).

  The wet component of the tropospheric propagation delay (the zenith
  wet delay, ZWD) is highly variable and cannot be predicted
  accurately.  If no tabulated ZWD information is available the effect
  is neglected, otherwise tabulated data may be used. With a typical
  excess zenith path length of 100--400~mm, error is incurred at the
  level of approximately 1.5\,ns.

\subsection{Einstein delay}\label{sec:einstein}

The Einstein delay \cite{dd86} quantifies the change in arrival times
due to variations in clocks at the observatory and the SSB due to
changes in the gravitational potential of the Earth and the Earth's
motion. IAU resolution A4 (1991) recommends the use of barycentric
coordinate time (TCB) which differs from TT both in mean rate and in
periodic and quasi-periodic terms.  By default, this is the coordinate
time in which arrival times are specified in \textsc{tempo2}.  Prior
to the definition of TCB, the recommended barycentric coordinate time
was Barycentric Dynamical time (TDB) which was implemented in
\textsc{tempo1}. In addition to being physically unrealisable
(Standish 1998)\nocite {sta98c}, TDB values are not physical
coordinate times, but rather values of a variable related to time by a
dimensionless scale factor \cite{kli05}. If these values are taken as
barycentric coordinate times of arrival, as has been common practice
in the past, then the scaling factor is effectively transferred from
the value to the units. Therefore, although site arrival times are
referred to TT, which is defined in terms of the SI second, TDB
barycentric arrival "time" intervals, and in fact, the numerical
values of all parameters inferred with pulsar timing on the basis of
TDB TOAs are effectively measured in units that differ subtly from
their SI counterparts.

As a result, all catalogued parameters measured using TDB TOAs (e.g.  
those from \textsc{tempo1}) must be multiplied by 
\begin{equation}
 K = 1 + (1.55051979154\times10^{-8} \pm 3\times10^{-17})
\end{equation}
\cite{if99}.  This can be a large effect and timing models created
using \textsc{tempo1} need to be modified before being used by
\textsc{tempo2}.  The n'th frequency derivative scales as $K^{-(n+1)}$
and the orbital period and semi-major axis all scale as $K$. The
epochs of periastron, period, position and dispersion measure all
scale as $K$ in their offset from the common epoch of Modified Julian
Day 43144.0003725 \cite{if99}.  For instance, the modification in
pulse frequency produces a slope of 0.5\,s\,yr$^{-1}$ in the timing
residuals which, for millisecond pulsars, will lead to phase coherence
being lost over even short data spans.  Using the \textsc{transform}
plug-in, \textsc{tempo2} provides an interface that can be used to
convert old parameters into the new system.  We also emphasise that,
because of the significant differences between the TDB and TCB, for
all published timing models the coordinate frame used must be clearly
specified.

 \subsection{Roemer delay}\label{sec:prop}

  The Roemer delay is the vacuum light travel time between the pulse
  arriving at the observatory and the equivalent arrival time at the
  SSB.  In \textsc{tempo2} this is calculated by determining the time
  delay between a pulse arriving at the observatory and at the Earth's
  centre and, with the aid of a Solar System ephemeris, from the
  Earth's centre to the SSB.

  The coordinates of the pulsar are known, either from telescope
  pointing, interferometry or pulsar timing, and are normally measured
  in the International Celestial Reference System (ICRS)\footnote{In
  the case of positions obtained by pulsar timing, this is only true
  if the reference frame of the Solar System ephemeris is tied to the
  ICRS, e.g. by using the DE405 planetary ephemeris. The DE200
  ephemeris is offset from the ICRS by $\sim$14 mas \cite{fcf+94},
  yielding a potentially significant error in the transfer to the
  geocentre if such a position is used; see Paper~II.}. The required
  transformation between the ICRS and the International Terrestrial
  Reference Frame (ITRF), within which observatory positions are
  determined, depends on precession, nutation, polar motion and Earth
  rotation.  The worst-case timing offset resulting from a positional
  error of $\Delta\theta$ is given by $\Delta\theta R_\oplus / c$,
  i.e.

  \begin{equation}
   \frac{\Delta t}{1\;\mathrm{ns}} \simeq 
    \frac{\Delta\theta}{9.7\;\mathrm{mas}} .
  \end{equation}

  \begin{figure*}
   \begin{center}
    \includegraphics[angle=-90,width=7cm]{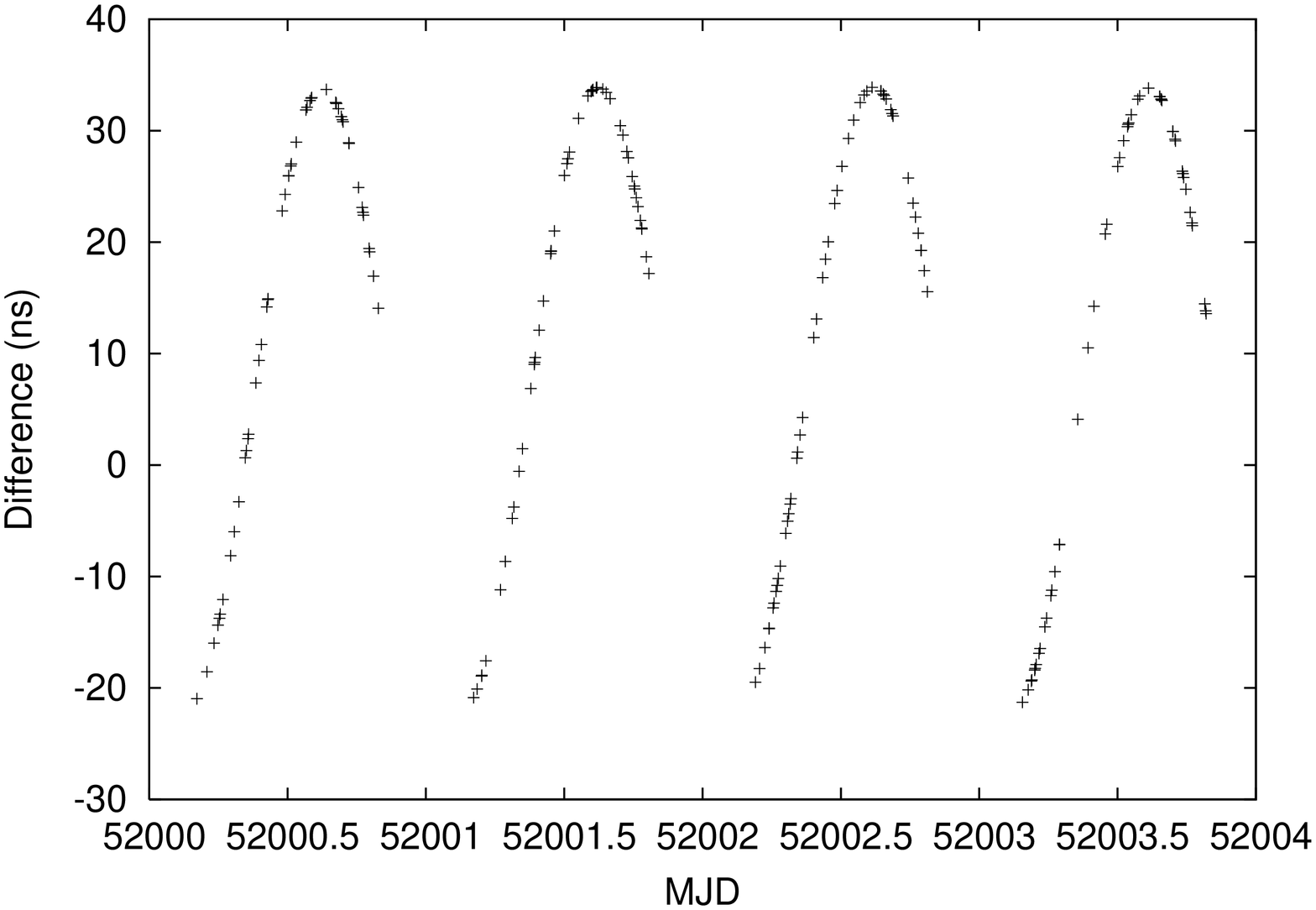}
    \includegraphics[angle=-90,width=7cm]{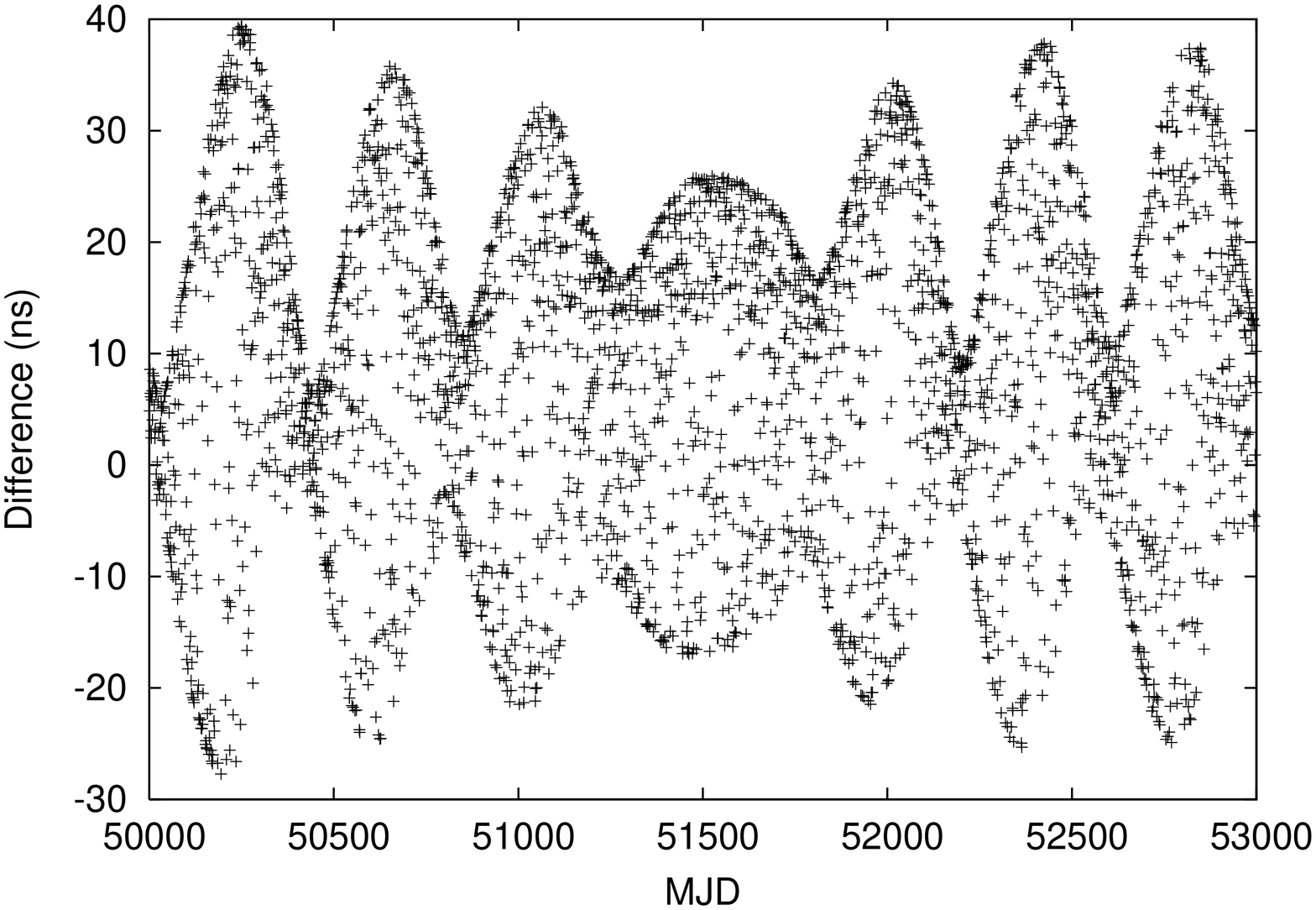}
   \end{center}
  \caption{Differences in the Solar System Roemer delay computed using
   current IAU precession-nutation models and including polar motion,
   versus the algorithm of \textsc{tempo1} for simulated TOAs from
   PSR~J1022$+$1001. The diurnal timing term is shown in the leftmost
   plot which is modulated by the yearly and 435-d periodicities of
   the polar motion (right) which, in turn, beat with a $\sim$6-year
   period.}\label{fg:polarmotion}
 \end{figure*}

  Through its omission of polar motion amounting to up to $\pm$ 300
  mas (corresponding to $\pm$30\,ns) and also through the use of IAU
  1976 precession \cite{llfm77} and IAU 1980 nutation \cite{sei82}
  models which are in error at the 50 mas (5\,ns) level, the
  \textsc{tempo1} software introduces errors in the timing model that
  are significant at contemporary levels of timing precision. In
  \textsc{tempo2}, polar motion is corrected using the values
  published in the C04 series of Earth Orientation Parameters (EOP) of
  the International Earth Rotation Service (IERS). The IERS also
  provides the difference between the observed precession and nutation
  and that predicted by the IAU 1976 and 1980 models. However,
  following the recommendations of IAU Resolutions adopted at the 24th
  General Assembly, we adopt the IAU 2000 precession-nutation model
  which provides sufficiently accurate predictions.  Specifically,
  \textsc{tempo2} uses the truncated 2000B model \cite{ml03} which is
  accurate to 1 mas (0.1\,ns).  Figure~\ref{fg:polarmotion} shows the
  differences between the Solar System Roemer delay computed using
  \textsc{tempo1} and \textsc{tempo2} using simulated observations of
  PSR~J1022$+$1001. Differences in the model (mainly due to polar
  motion) introduce an error in the assumed observatory position,
  which appears as a diurnal timing term which is modulated by the
  yearly and 435-d periodicities of the polar motion.

  The third component in the transformation of the pulsar position to
  the ITRS is the Earth rotation angle which is a linear function of
  the time scale known as UT1. This is computed by \textsc{tempo2}
  using the offset between UTC and UT1 as provided in the C04 EOP
  series. 

  \begin{figure*}
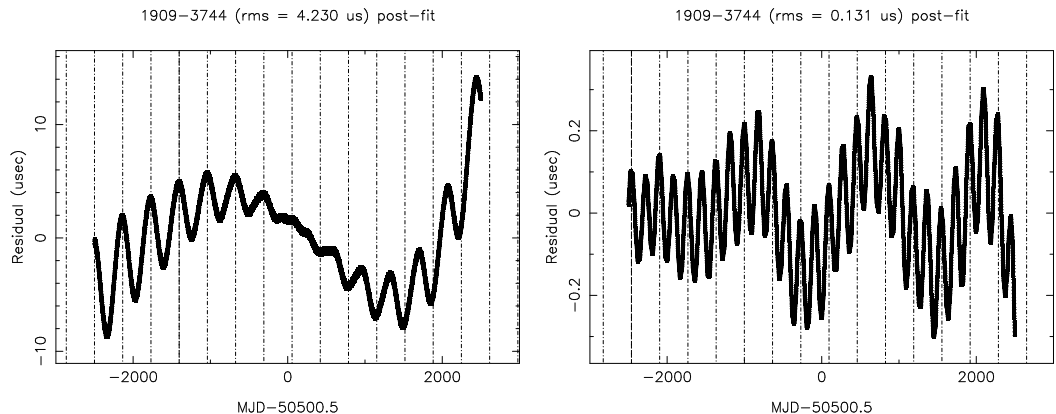

   \begin{center}
    \includegraphics[angle=-90,width=7cm]{compareDE200_DE405.ps}
    \includegraphics[angle=-90,width=7cm]{compareDE200_DE405_2.ps}
   \end{center}
  \caption{Comparison between timing residuals obtained using the
  DE200 Solar System ephemeris and the DE405 ephemeris.  In the
  left plot, terms corresponding to a pulsar position error,
  spin-frequency and its first derivative have been subtracted. In the
  right plot, terms corresponding to the above and higher frequency
  derivatives, orbital period, epoch of periastron, proper motion and
  parallax have also been removed. The vertical lines are spaced at
  1\,year intervals. These plots were created using the \textsc{plk}
  plug-in for \textsc{tempo2}.}\label{fg:de200_de405}
 \end{figure*}

  The choice of Solar System ephemeris for determining the position of
  the SSB with respect to the Earth can have significant effects on
  the calculated timing residuals.  Until recently the JPL DE200 model
  \cite{sta90}, which is based upon the dynamical equator and equinox
  of J2000, was the most widely used. More recently, the JPL DE405
  model has been
  developed\footnote{ftp://ssd.jpl.nasa.gov/pub/eph/export/DE405/de405.iom/}
  which, in contrast to the DE200 model, is aligned with the ICRS
  (Standish 1998)\nocite{sta98c}. The DE405 model includes the
  planets, the Earth's Moon and 300 asteroids. In the left panel of
  Figure~\ref{fg:de200_de405} we plot the difference between residuals
  obtained using the DE405 and DE200 models after the subtraction of
  an annual sinusoid (corresponding to a position error) and quadratic
  term (corresponding to the spin-frequency and its first derivative)
  for simulated observations of PSR~J1909$-$3744. The right panel
  contains the timing residuals after fitting for five frequency
  derivative terms, the orbital period, epoch of periastron, proper
  motion and parallax.  The use of the DE200 model leads to an
  incorrect measurement of the proper motion in right ascension by
  $-0.2775(3)$\,mas\,yr$^{-1}$, in declination by
  $-0.037(1)$\,mas\,yr$^{-1}$ and parallax by $-0.045(5)$\,mas over
  this simulated, regularly sampled data span of 14\,years.  Splaver
  et al. (2005)\nocite{sns+05} also reported significant deviations
  between residuals for PSR~J1022$+$1001 using these Solar System
  ephemerides during the years 1998-1999 which they explained by the
  new ephemeris incorporating improved measurements of the outer
  planet masses.  Although \textsc{tempo2} can access any of the JPL
  planetary ephemerides, we currently recommend that the DE405 model
  be used for any high precision analysis of pulse arrival times.
 
  \subsubsection{Shapiro delay}\label{sec:shapiro}

\begin{figure}
   \begin{center}
    \includegraphics[angle=-90,width=7cm]{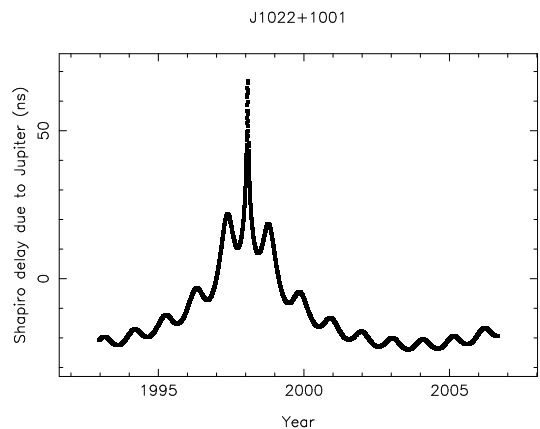}
   \end{center}
  \caption{The additional time delay from the Shapiro delay due to
  Jupiter for PSR~J1022$+$1001.}\label{fg:1022Jupiter}
 \end{figure}

\begin{figure*}
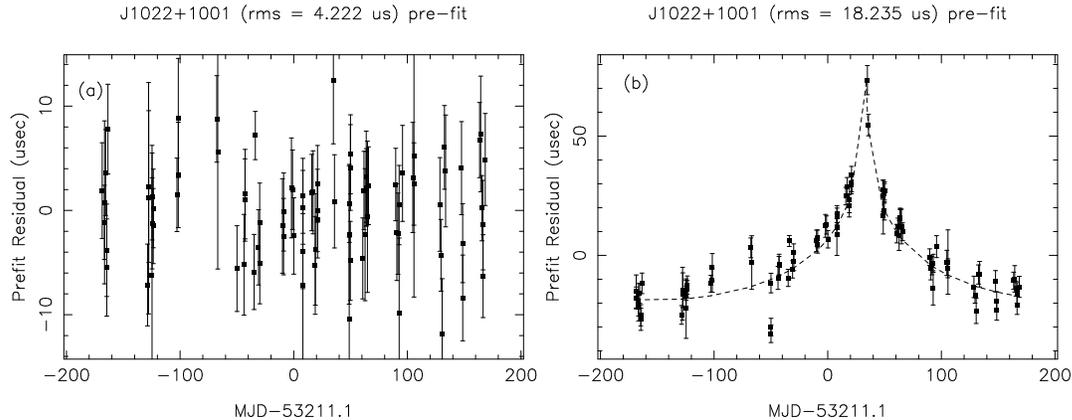

   \begin{center}
    \includegraphics[angle=-90,width=7cm]{shapiro2.ps}
    \includegraphics[angle=-90,width=7cm]{shapiro.ps}
   \end{center}
  \caption{The timing residuals, in $\mu s$, for PSR~J1022$+$1001, a)
  after fitting for the pulsar's parameters and b) without removal of the Solar
  System Shapiro delay. This plot was created using the
  \textsc{plk} plug-in for \textsc{tempo2} (note, the original plk
  plotting package incorrectly plotted the uncertainties on the
  residuals; the errors were a factor of two too
  small).}\label{fg:shapiro}
 \end{figure*}

  To make an accurate determination of the arrival time at the
  barycentre it is also necessary to include the Shapiro delay due to
  Solar System objects (most notably the Sun) which accounts for the
  time delay caused by the passage of the pulse through large
  gravitational fields \cite{sha64}. Table \ref{tb:corr} shows the
  maximum variation in Shapiro delay for a selection of Solar System
  bodies. \textsc{Tempo2} includes all bodies for which the maximum
  variation is greater than 0.1\,ns.  Figure~\ref{fg:1022Jupiter}
  shows the variations in the Shapiro delay due to Jupiter for
  PSR~J1022$+$1001\footnote{The Shapiro delay as characterised by
  Damour \& Dereulle (1986) can be negative. However, as the
  zero-order time of arrival of the pulses is arbitrary, a constant
  offset can be added to the Shapiro delay calculation.}. The effect
  of the Shapiro delay due to the Sun can be clearly seen in the
  observations of PSR~J1022$+$1001 which has an ecliptic latitude of
  $-0.06^\circ$. In Figure~\ref{fg:shapiro}a we plot the pulse timing
  residuals after fitting for the pulsar's parameters and the Solar
  System Shapiro delay.  Figure~\ref{fg:shapiro}b shows the resulting
  timing residuals if the best-fit parameters are used, but the Solar
  System Shapiro delay is not calculated when forming the barycentric
  arrival times. On MJD~53143, this pulsar passed within 5$^\circ$ of
  Jupiter.  However, the additional Shapiro delay due to Jupiter is
  not detectable with our current data.

 \subsection{Frequency-dependent parameters}\label{sec:dm}

  \begin{table*}
   \caption{\textsc{Tempo2} parameters relevant to frequency-dependent
   offsets.}\label{tb:dms}
   \begin{tabular}{llll} \hline
   Parameter & Description & Symbol \\ \hline
   DM, DM1 \ldots & The dispersion measure and its derivatives & ${\rm DM}$,
   $\dot{\rm DM}$ \ldots \\
   DMEPOCH    & The epoch of the disperison measure (MJD) & $t_{\rm D}$ \\
   FDDI       & Index for frequency dependent delay & $\zeta$ \\
   FDDC       & Scale for frequency dependent delay & $k_{\rm f}$ \\
   \hline\end{tabular}
  \end{table*}

   \textsc{Tempo2} provides the ability to fit for delays which are
   dependent upon the observing frequency; see Table~\ref{tb:dms}.
   For instance, dispersion measure (DM) delays are $\propto f^{-2}$
   whereas delays caused by refractive and diffractive effects are
   $\propto f^{-4}$ (e.g. Foster \& Cordes
   1990)\nocite{fc90}. \textsc{Tempo2} allows fitting for a parameter
   that is $\propto f^{-\zeta}$ where $\zeta$ is defined by the user
   and is not restricted to integral values.  We emphasise that in
   order to obtain absolute values for these frequency-dependent terms
   it is necessary to obtain TOAs using aligned standard templates.
   In practice, true absolute alignment is impossible because of
   profile shape evolution with frequency, so frequency-dependent
   parameters are always relative at some level.

   Although DM values are commonly published, the directly measurable
   parameter from pulsar timing observations is $D$, the dispersion
   constant, where

   \begin{eqnarray}  
    {\rm DM} &=& D/k_{\rm D} {\rm .}
   \end{eqnarray}
   If the effect of ions and magnetic fields in the interstellar
   medium are ignored, then
   \begin{eqnarray}
    k_{\rm D} &=& \frac{e^2}{\pi^2 m_e c}{\rm .}
   \end{eqnarray}
   However, ions and magnetic fields introduce a rather uncertain
   correction of order a part in 10$^5$ \cite{spi62}, comparable to
   the uncertainty in some measured DM values (e.g. Phillips \&
   Wolszczan 1992\nocite{pw92}). Consequently, both \textsc{tempo1}
   and \textsc{tempo2} adopt a value of $k_{\rm D} \equiv 2.410 \times
   10^{-16}$\,cm$^{-3}$\,pc\,s (Manchester \& Taylor
   1977)\nocite{mt77}.  It is also possible, in \textsc{tempo2}, to
   set $k_{\rm D} = 1$ in order to measure the dispersion constant.
 
\begin{figure}
    \includegraphics[angle=-90,width=7cm]{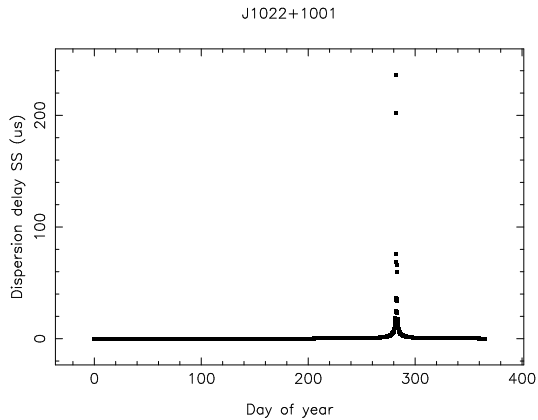}
    \caption{The extra time-delay automatically added by
    \textsc{tempo2} to model the interplanetary medium for
    PSR~J1022$+$1001 at an observing frequency of 1400\,MHz.  This
    figure was produced using the \textsc{fake} and \textsc{delays}
    plugins for \textsc{tempo2}.}
    \label{fg:1022_ss_dm}
   \end{figure}

   Another dispersive component occurs in the Solar System.  The
   interplanetary medium is dominated by the Solar wind and is
   approximated in \textsc{tempo2} with the electron density
   decreasing as an inverse square law from the centre of the Sun
   (full details are provided in Paper~II) with $n_0$ being the
   electron density at the Earth.  \textsc{Tempo1} uses $n_0 =
   9.961$\,cm$^{-3}$. However, by default, \textsc{tempo2} uses a value
   of $n_0 = 4$\,cm$^{-3}$ which is more consistent with recent
   measurements \cite{immh98}. Figure~\ref{fg:1022_ss_dm} shows the
   extra time delay added by \textsc{tempo2} for simulated
   observations of PSR~J1022$+$1001. As discussed in Paper~II, this
   estimation of the extra time delay is poor as the true electron
   density can vary dramatically.  We therefore recommend that, for
   high-precision timing, \textsc{tempo2} be provided with
   multiple frequency observations which allow the determination of
   the actual DM for each observation.

   As an example, the PPTA uses a dual-band receiver at 10 and 50\,cm.
   The DM at any instant can therefore be measured to a precision of
   up to $1\times10^{-4}$\,cm$^{-3}$pc if the difference between
   observations at the two frequencies can be measured to 1\,$\mu s$.
   \textsc{Tempo2} can be run in a mode where simultaneous (or
   contemporaneous) observations at multiple frequencies are used to
   determine the current DM and to use that value in subsequent
   calculations.  For times when there are no multi-frequency
   observations available, the DM can be estimated from a polynomial
   fit to measured DM values before and after the observation.

 \subsection{Shklovskii effect and radial motion}\label{sec:vacuum}

  Pulsar timing measurements are affected by the secular motion of the
  pulsar relative the SSB. In the past the secular terms involving
  this motion have been omitted from timing models, because they can
  be absorbed in alterations of other parameters. The four largest
  effects are the radial velocity (affecting most spin and binary
  parameters; Damour \& Deruelle 1986\nocite{dd86}), the Shklovskii
  effect and radial acceleration (affecting the spin and orbital
  period derivatives; Shklovskii 1970, Damour \& Taylor
  1991\nocite{shk70}\nocite{dt91}) and the mixing of radial velocity
  into the Shklovskii term (affecting the spin period second
  derivative; van Straten 2003\nocite{van03a}). In contrast to
  \textsc{tempo1}, \textsc{tempo2} takes the approach that these terms
  can be included in the timing model as long as steps are taken to
  ensure the model is sufficiently constrained. In this way, one may
  take into account what is known about the secular motion and
  distance (via, for example, its appearance in the annual proper
  motion and parallax terms) to provide correct measurements of the
  spin and orbital parameters, rather than measuring incorrect values
  and attempting to correct them post facto (e.g. Damour \& Taylor
  1991\nocite{dt91}). Conversely, if one may safely assume that one of
  the affected spin or orbital parameters is zero, it may be held
  fixed at this value in order to obtain a direct measurement of the
  distance or velocity, rather than measuring incorrect spin and
  orbital parameter values and using these to infer the motion and
  distance indirectly (e.g. Bell \& Bailes 1996, van Straten
  2003\nocite{bb96}\nocite{van03}).

\section{Fitting routines}\label{sec:fit}

  \textsc{Tempo2} uses the derived time of emission and a given timing
  model to form the i'th pre-fit timing residual:
  \begin{equation}
   R_i = \frac{\phi_i - N_i}{\nu}.
  \end{equation}
 $\phi_i$ describes the time evolution of the pulse phase based on the
 model pulse frequency ($\nu$) and its derivatives in addition to any
 glitch parameters. $N_i$ is the nearest integer to $\phi_i$.
 Paper~II contains details for calculating $\phi_i$.


 Terms corresponding to offsets in model
 parameters are fitted to these residuals in order to improve the
 measurement of these parameters.  By default, the entire procedure is
 repeated using the post-fit timing model in order to produce accurate
 post-fit barycentric arrival times and residuals.  This is in
 contrast to \textsc{tempo1} which only obtains the barycentric
 arrival times once and predicts the expected post-fit timing
 residuals. This entire process often needs to be iterated until
 convergence is reached as the offsets made to model parameters are
 based on a linearised approximation to the effects on the timing
 model (e.g. Damour \& Deruelle 1986\nocite{dd86}).

\begin{figure}
   \begin{center}
    \includegraphics[angle=-90,width=7cm]{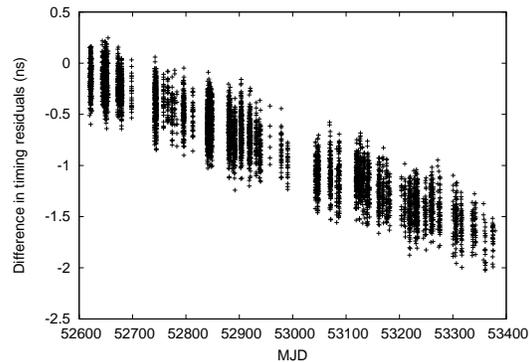}
   \end{center}
  \caption{Comparison between \textsc{tempo1} and \textsc{tempo2}
  timing residuals for the PSR~J0437$-$4715 observations when
  \textsc{tempo2} is emulating \textsc{tempo1}. No clock corrections
  were applied to the TOAs.}\label{fg:diff}
 \end{figure}

 The fitting routines in \textsc{tempo2} are based on a linear
 singular-value decomposition, weighted least-squares algorithm (for
 example, Press et al. 1992\nocite{ptvf92})\footnote{It is possible
 with plug-in capabilities to use a non-linear fitting algorithm, but
 this is not necessary for the routines described in this paper; more
 details will be provided in Paper~III.} where
 \begin{equation}
  \chi^2 = \sum_{i=1}^{N} \left(\frac{R_i}{\sigma_i}\right)^2
 \end{equation}
 is minimised.  $N$ is the number of observations and $\sigma_i = 1$
 for unweighted fits or set to the TOA uncertainty for the i'th
 observation for a standard $\chi^2$-minimisation.  If specified as
 Modified Julian dates, measured TOAs need to be accurate to better
 than 19 significant figures for 1\,ns timing precision and
 spin frequencies are now routinely measured to 16 significant
 figures.  In \textsc{tempo2} all parameters are stored and all
 calculations are carried out with ``long double'' precision that
 typically provides 12\,bytes of storage (allowing 18 significant
 digits) on PC-based systems and 16\,bytes (33 significant digits) on
 most other systems.  In Figure~\ref{fg:diff} we plot the difference
 between the pre-fit timing residuals for PSR~J0437$-$4715 obtained
 using \textsc{tempo1} and \textsc{tempo2}. The observed trend that
 covers $\sim$2\,ns over 2.1 years of observing is due to
 \textsc{tempo1} not storing the pulse frequency with enough
 significant figures and will lead to \textsc{tempo1} introducing
 systematic effects in the timing residuals over long time spans.

\begin{figure}
    \begin{center}\includegraphics[angle=-90,width=7cm]{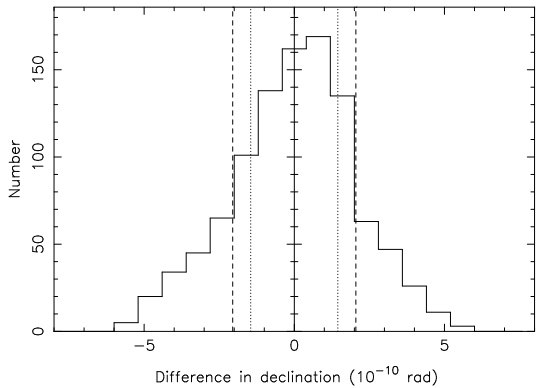}
    \caption{Histogram of the declination parameter obtained using the
    bootstrap technique for PSR~J1909$-$3744.  The dashed lines
    indicate the 1$\sigma$ uncertainties obtained using this method
    and the dotted lines give the 1$\sigma$ uncertainties measured
    using the standard least-squares-fitting routine.}\label{fg:histBS}\end{center}

%
%
   \end{figure}

 It is also useful to compare the formal uncertainties described above
 with those obtained using a ``bootstrapping'' method (see, for
 instance, Wall \& Jenkins 2003\nocite{wj03}) which can produce more
 realistic parameter values and uncertainties when significant
 correlations between parameters are present. The bootstrapping method
 implemented in \textsc{tempo2} estimates the uncertainty on a
 parameter by 1) randomly selecting observations to produce a new
 dataset of the same length as the original (the observations are
 selected with replacement; i.e. in the new dataset some of the
 original observations will be omitted while others will be
 replicated), 2) recalculating the parameter and 3) repeating as many
 times as possible\footnote{Note: the bootstrapping method implemented
 in \textsc{tempo1} does not re-fit for the parameters after randomly
 selecting observations and is therefore not a true bootstrapping
 technique.}. The distribution of these parameters provides an
 estimate of the uncertainty (obtained from the standard deviation of
 the distribution) on the value of the parameter (taken as the mean of
 the distribution).  In Table~\ref{tb:1909_bs}, we compare the values
 and uncertainties on the fitted parameters using the formal
 least-squares fitting and the bootstrap technique with 1024
 iterations for PSR~J1909$-$3744.  For this pulsar, with residuals
 that are not dominated by timing noise, the measured uncertainties
 are typically $\sim 1.2$ times larger with the bootstrap technique
 than with the least-squares method. A histogram of the fitted
 declination is shown in Figure~\ref{fg:histBS}.

 \begin{table*}
  \caption{Comparison between standard least-squares (LS) parameters and
  uncertainties with those obtained using a bootstrapping (BS)
  technique for PSR~J1909$-$3744.}\label{tb:1909_bs}
  \begin{tabular}{llllll}\hline
   Parameter & Value ($V_{LS}$) & Error ($E_{LS}$) & Value ($V_{BS}$)
   & Error ($E_{BS}$) & $E_{BS}/E_{BS}$ \\ \hline
   Right ascension (rad) & 5.016908214879 & 3.5$\times 10^{-11}$ &
               5.016908214880 & 4.3$\times 10^{-11}$ & 1.2 \\
   Declination (rad)& $-$0.65863987098 & 1.4$\times 10^{-10}$ &
               $-$0.65863987100 & 2.1$\times 10^{-10}$ & 1.5 \\ 
   Pulse frequency (Hz) & 339.31568762926 & 1.9$\times 10^{-12}$ &
                         339.31568762926 & 2.4$\times 10^{-12}$ & 1.3 \\
   Frequency derivative (s$^{-2}$) & $-$1.614873$\times 10^{-15}$ &
   2.4$\times 10^{-20}$ &            $-$1.614878$\times 10^{-15}$ &
   2.9$\times 10^{-20}$ & 1.2 \\ \\
   Orbital period (d) & 1.533449474188 & 1.1$\times 10^{-11}$ &
                        1.533449474191 & 1.4$\times 10^{-11}$ & 1.3 \\
   Projected semi-major axis (lt-s) & 1.897991295 & 9.9$\times 10^{-9}$ &
                            1.897991295 & 1.2$\times 10^{-8}$ & 1.2 \\
   Epoch of periastron (MJD) & 52053.452 & 0.021 & 
                               52053.443 & 0.020 & 0.95 \\
   Eccentricity & 1.186$\times 10^{-7}$ & 9.5$\times 10^{-9}$ &
                  1.187$\times 10^{-7}$ & 9.6$\times 10^{-9}$ & 1.0 \\
  \hline\end{tabular}
  
 \end{table*}

\begin{figure}
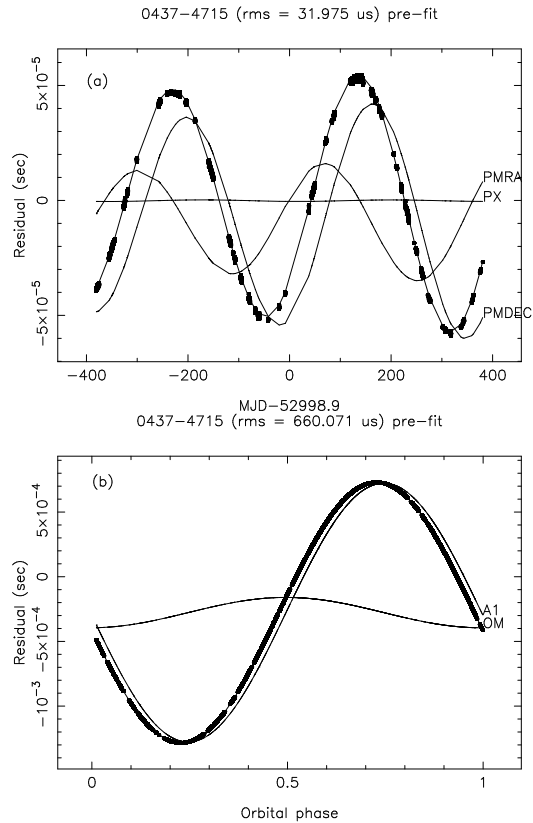

   \begin{center}
    \includegraphics[angle=-90,width=7cm]{component1.ps}
    \includegraphics[angle=-90,width=7cm]{components2.ps}
   \end{center}
  \caption{The components of a fit to update a pulsar's a) proper
  motion and parallax and b) longitude of periastron and projected
  semi-major axis of orbit (assuming that these are the only
  parameters being updated). The line through the measured residuals
  indicates the sum of the model components.  These plots were created
  using the \textsc{plk} plug-in for \textsc{tempo2}.}\label{fg:model}
 \end{figure}
\begin{figure}
   \begin{center}
    \includegraphics[angle=-90,width=7cm]{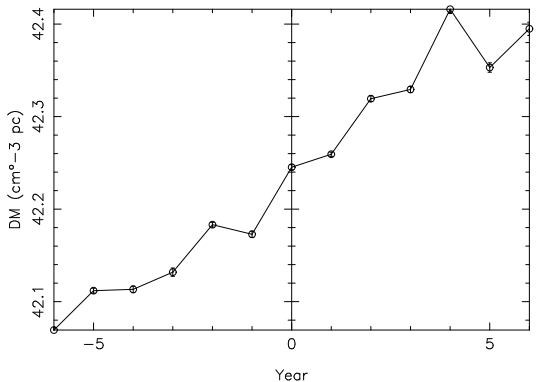}
   \end{center}
  \caption{The dispersion measure variation for PSR~B0458+46 obtained by fitting
  for dispersion measure in 1\,year sections of Jodrell Bank
  observatory data.  This plot was produced using the
  \textsc{stridefit} plug-in for \textsc{tempo2}.}\label{fg:dmdot}
 \end{figure}

 Various interfaces exist that allow the user to study the actual fits
 being applied to the pre-fit timing residuals.  For instance, in
 Figure~\ref{fg:model}a we plot the components of the fit when
 improving a pulsar's proper motion and parallax.  In
 Figure~\ref{fg:model}b we demonstrate the effects of poorly estimated
 binary parameters.  The \textsc{tempo2} software also provides the
 ability to fit one or more of the timing model parameters over short
 adjacent subsets of the data.  For example, this allows the user to
 analyse dispersion measure variations (Figure~\ref{fg:dmdot}), search
 for glitch events and to confirm proper motions.

  It is often difficult, with sparse observations, to obtain an
  accurate timing model.  This is often solved by making further
  observations of the pulsar, but with the \textsc{gorilla} plug-in,
  \textsc{tempo2} provides an alternative method.  For example, 35
  observations of PSR~J0857$-$4424 spread over seven years were
  obtained at the Parkes observatory, but a solution producing
  phase-connected timing residuals could not be
  obtained. \textsc{Gorilla} provides a brute-force fitting technique
  that obtains the pre-fit timing residuals over many millions of
  combinations of spin-frequency and its derivative within specified
  ranges.  This method found the correct solution over a 1-year
  section of the PSR~J0857$-$4424 data and this fit was extrapolated
  to phase-connect the entire seven years of observations without the
  necessity for further observations.

%

 


\section{The post-fit parameters and their errors}

 \subsection{Pulsar spin parameters}

  \textsc{Tempo2} provides the ability to fit for the pulsar's pulse
  frequency and an arbitrary number of spin-frequency derivatives.
  The full set of possible spin parameters available are listed in
  Table~\ref{tb:spin}.  Except for the very youngest pulsars, the
  frequency second and higher derivatives are not believed to
  represent the secular spin-down of the pulsar, but rather timing
  irregularities known as timing noise (e.g. Hobbs et
  al. 2004)\nocite{hlk+04}.  For pulsars whose timing residuals are
  dominated by timing noise it is not possible to determine accurate
  positions, proper motions or dispersion measures without
  ``whitening'' the data while fitting the timing model.
  Traditionally, this whitening procedure has been carried out by
  fitting multiple spin-frequency derivatives until the resulting
  post-fit residuals are free of systematic structure.  As described
  by Hobbs et al. (2004)\nocite{hlk+04} this whitening technique is
  limited because 1) only low-frequency timing noise can be modelled
  without affecting the higher-frequency signatures of position errors
  and proper motions and 2) such whitening is limited to polynomials
  of order 12 to prevent floating-point overflows. Hobbs et
  al. (2004)\nocite{hlk+04} described a new method, based on the
  fitting of harmonically related sinusoids, that produces superior
  results in many cases.  This sinusoidal fitting technique has been
  implemented into \textsc{tempo2}.  In brief, \textsc{tempo2} obtains
  from the parameter file a fundamental frequency ($\omega$) and the
  amplitudes of $n_H$ harmonically related sinusoids. If the
  fundamental frequency is not provided then it is derived from the
  observation span ($T_{\rm span}$) as
  \begin{equation}
   \omega = \frac{2 \pi}{T_{\rm span}(1+4/n_H)}.
  \end{equation} 
  \textsc{Tempo2} subsequently subtracts
  \begin{equation}
   \Delta R = \sum_{k=1}^{n_H}a_k \sin (k \omega \Delta t) + b_k \cos
   (k \omega \Delta t)
  \end{equation}
  from the timing residuals where $\Delta t$ represents the difference
  between an arrival time in the pulsar reference frame and the
  specified period epoch. \textsc{Tempo2} can also fit for the
  coefficients $a_i$ and $b_i$ and output these parameters as part of
  a new timing model. In contrast to the technique described by Hobbs
  et al. (2004), the default method implemented by \textsc{tempo2}
  simultaneously fits for the pulsar parameters and for the sinusoids,
  i.e. this is not a pre-whitening technique.

  \begin{table*}
   \caption{The spin, glitch and whitening parameters included in \textsc{tempo2}.}\label{tb:spin}
   \begin{tabular}{llll} \hline
   Parameter & Description & Symbol \\ \hline
   F0, F1 \ldots & The pulse-frequency and its derivatives & $\nu$,
   $\dot{\nu}$ \ldots \\
   PEPOCH    & The epoch of the pulse-frequency measurement & $t_P$ \\
   \\
   GLEP\_k    & Glitch epoch (MJD) & $t_g$ \\
   GLPH\_k    & Glitch phase increment & $\Delta \phi_g$ \\
   GLF0\_k    & Glitch permanent pulse frequency increment (Hz) & $\Delta \nu_g$\\
   GLF1\_k    & Glitch permanent frequency derivative increment
   (s$^{-2}$) & $\Delta \dot{\nu}_g$ \\
   GLF0D\_k   & Glitch decaying pulse frequency increment (Hz) & $\Delta \nu_{d}$\\
   GLTD\_k    & Glitch decay time constant (d) & $\tau_d$ \\
   \\
   WAVE\_OM    & Fundamental frequency of sinusoids for whitening (Hz) & $\omega$ \\ 
   WAVE\_k     & Amplitude of the sine and cosine terms for the k'th sinusoids & $a_k$,$b_k$\\
   \hline\end{tabular}
  \end{table*}

\begin{figure*}
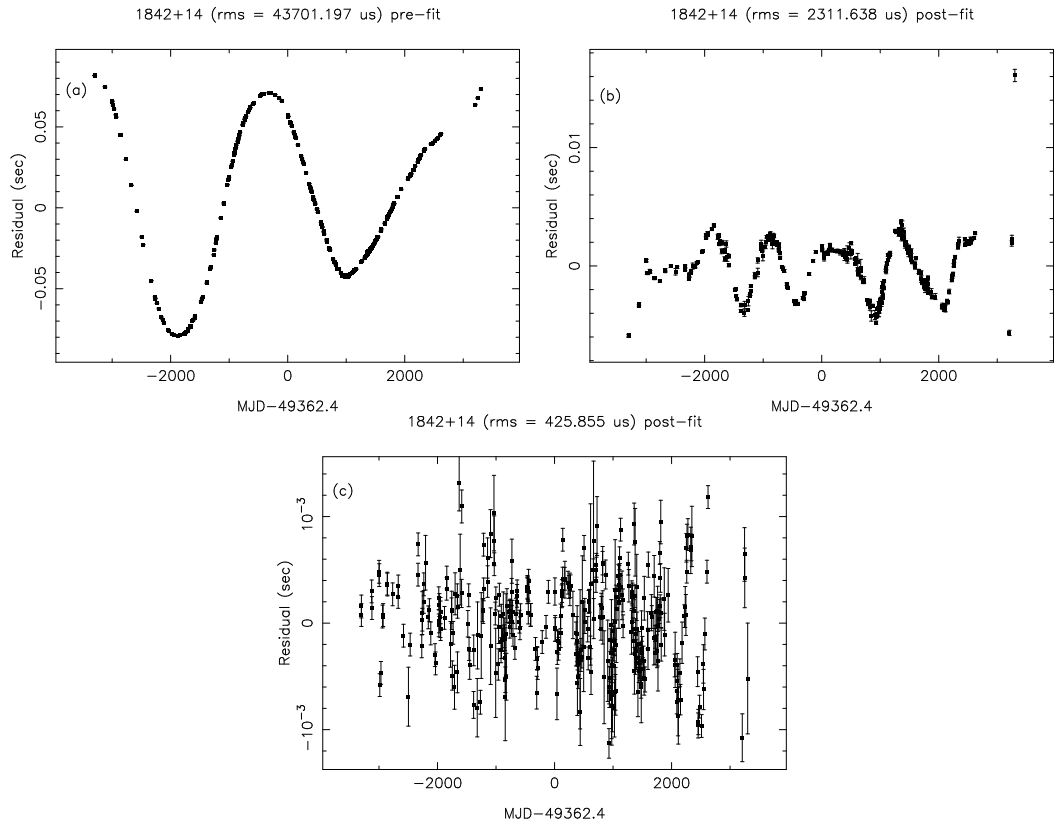

   \begin{center}
    \includegraphics[angle=-90,width=7cm]{pre.ps}
    \includegraphics[angle=-90,width=7cm]{white2.ps}
    \includegraphics[angle=-90,width=7cm]{white1.ps}
   \end{center}
  \caption{a) The pre-fit timing residuals for PSR~B1842$+$14 obtained
  from the Jodrell data archive. b) Whitening the timing residuals
  using 11 polynomial coefficients and c) whitening the timing
  residuals using sinusoids.}\label{fg:white}
 \end{figure*}

  An example of such ``whitening'' of the data is shown in
  Figure~\ref{fg:white}. The timing residuals of PSR~B1842$+$14
  obtained from the Jodrell Bank data archive are typical of those
  seen for non-recycled pulsars over many years of observation.  They
  are dominated by quasi-periodic structures that are well modelled
  using the sinusoidal modelling, but not by high-order polynomial
  terms.
  
  The pulse frequencies of many young pulsars have been observed to
  increase suddenly during a glitch event.  An individual glitch can
  be characterised by the epoch of the glitch ($t_g$), the phase
  increment at the glitch ($\Delta \phi_g$), a permanent pulse
  frequency increment ($\Delta \nu_g$), a permanent frequency
  derivative increment ($\Delta \dot{\nu}_g$), a decaying pulse
  frequency increment ($\Delta \nu_{d}$) and the decay time constant
  ($\tau_d$).  The increase in pulse phase due to a glitch event at
  time $t_g$ is given by
  
  \begin{equation}
   \phi_g = \Delta \phi_g + \Delta t \Delta \nu_g  + \frac{1}{2}\Delta
   \dot{\nu}_g (\Delta t)^2 + \tau_d\Delta \nu_{d}(1 - e^{-\Delta t/\tau_d})
  \end{equation}
  where $\Delta t = t-t_g$ is the time since the glitch event.
  \textsc{Tempo2} allows the user to fit for an arbitrary number of
  glitch events in a single dataset and provides various plug-ins to
  aid in their detection.

 \subsection{Astrometric parameters}

  \textsc{Tempo2} performs computations in the reference frame of the
  Solar System ephemeris which is approximately equatorial.  Since the
  location of the observatory is specified in the ITRS, and
  transformed to the ICRS according to the specifications of the IAU
  2000 resolutions, for the best possible accuracy a Solar System
  ephemeris that is aligned with the ICRS should be used. The JPL
  DE405 ephemeris is the most recent of the publically available JPL
  series and meets this criterion.  The \textsc{tempo2} equatorial
  astrometric parameters (``RA'', ``DEC'', etc.)  strictly refer to
  the ephemeris frame, which, even in the case of DE405, is only tied
  to the ICRS within a finite uncertainty. The ICRS itself is
  measurably offset from both the Fundamental Katalog 5 (FK5; Feissel
  \& Mignard 1998\nocite{fm98}) and the dynamical equator and equinox
  of J2000.0 \cite{hh04}. Seidelmann \& Kovalevsky (2002)\nocite{sk02}
  state that ICRS coordinates should be denoted ``right ascension''
  and ``declination'' with no further qualification. Since the
  coordinates measured using pulsar timing are relative to the
  coordinate frame of the chosen planetary ephemeris, the latter needs
  to be specified when quoting fitted positions. Owing to the known
  offset between J2000.0 and the ICRS we recommend that the common
  practise of labelling pulsar timing coordinates as ``J2000'' be
  discontinued.

  \textsc{Tempo2} also accepts astrometric parameters specified in the
  ecliptic frame.  For such parameters, \textsc{tempo2} transforms all
  vectors into the ecliptic coordinate system by rotating about the
  ``$x$-axis'' by the mean obliquity of the ecliptic, $\epsilon$, at
  the epoch J2000.0. By default, \textsc{tempo2} uses the current best
  estimate of the mean obliquity of the ecliptic, $\epsilon_{\rm
  DE405} = 84381.40578$\,arcsec \cite{hf04}. This value is derived
  from a harmonic decomposition of the DE405 Solar System ephemeris,
  and applies to the epoch J2000.0. This differs from the earlier
  estimate of 84381.412\,arcsec used by \textsc{tempo1}. Since timing
  pulsar positions are largely constrained by the Solar System Roemer
  delay, error ellipses tend to be aligned with ecliptic latitude or
  longitude, making this basis a convenient choice when one coordinate
  is constrained more strongly than the other. For instance, fitting
  for the proper motion in equatorial coordinates with the
  PSR~J1022$+$1001 dataset discussed in \S2 gives a proper motion in
  right ascension $\mu_\alpha = -148(55)$\,mas\,yr$^{-1}$ and in
  declination of $\mu_\delta = -335(143)$\,mas\,yr$^{-1}$ and the
  astrometric parameters are highly covariant in the fit.  Fitting
  using ecliptic coordinates provides a precise measurement of the
  proper motion in ecliptic longitude $\mu_\lambda =
  -16.1(2)$\,mas\,yr$^{-1}$ which can be used in 1-dimensional studies
  of pulsar velocities (e.g. Hobbs et al. 2005\nocite{hllk05}) and a
  much poorer determination in ecliptic latitude $\mu_\beta =
  -307(152)$\,mas\,yr$^{-1}$.  A full list of the astrometric
  parameters that can be used by \textsc{tempo2} is listed in
  Table~\ref{tb:astrometric}.

  \begin{table*}
   \caption{Astrometric parameters included in \textsc{tempo2}.}\label{tb:astrometric}
   \begin{tabular}{llll} \hline
   Parameter & Description & Symbol \\ \hline
   RA        & Right ascension of pulsar (hr min sec) &  $\alpha$  \\
   DEC       & Declination of pulsar (deg min sec) &  $\delta$ \\
   ELONG     & Ecliptic longitude of pulsar (deg) & $\lambda$ \\
   ELAT      & Ecliptic latitude of pulsar (deg) & $\beta$ \\
   PMRA      & Proper motion in right ascension (mas\,yr$^{-1}$) & $\mu_\alpha$ \\
   PMDEC     & Proper motion in declination (mas\,yr$^{-1}$) & $\mu_\delta$ \\
   PMELONG   & Proper motion in ecliptic longitude (mas\,yr$^{-1}$) & $\mu_\lambda$ \\
   PMELAT    & Proper motion in ecliptic latitude (mas\,yr$^{-1}$) & $\mu_\beta$ \\
   PX        & Parallax (mas) & $\Pi$ \\
   PMRV      & Radial proper motion & $\mu_\parallel$\\
   POSEPOCH  & Position epoch (MJD) & $t_{\rm pos}$\\
   \hline\end{tabular}
  \end{table*}

 \subsection{Binary parameters}\label{sec:binary}

  For pulsars in binary systems, \textsc{tempo2} includes terms that
  describe the pulsar's orbital motion.  Various timing models are
  available.  A full mathematical description of these models and
  their implementation in \textsc{tempo2} is provided in Paper~II.
  Here, we provide a summary.

\begin{figure}
   \begin{center}
    \includegraphics[angle=-90,width=7cm]{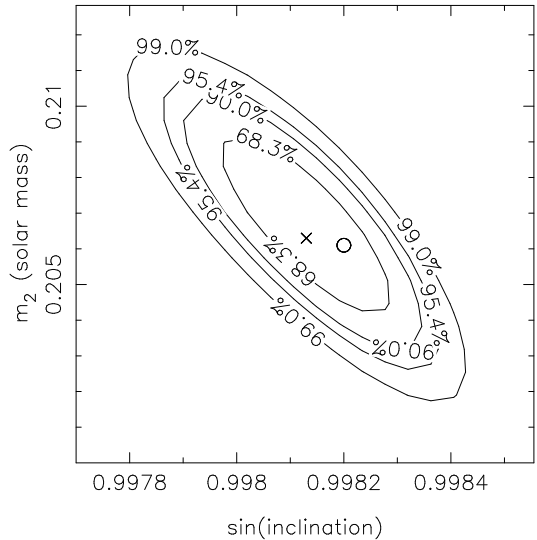}
   \end{center}
  \caption{$\chi^2$ contours in $m_2$-$\sin i$ parameter space for
  PSR~J1909$-$3744.  The circle indicates the most-likely values
  obtained using the `DD' model and the cross symbol for the `T2'
  model. This plot was produced using the \textsc{m2sini}
  plug-in.}\label{fg:1909_m2sini}
  \end{figure}

  The main binary model implemented in \textsc{tempo2} is referred to
  as the `T2' model and is based on the Damour \& Deruelle
  (1986)\nocite{dd86} model (`DD') implemented in \textsc{tempo1}. In
  contrast to the `DD' model which is designed for a pulsar and a
  single companion, the new `T2' model allows multiple binary
  companions.  Various other models were available to \textsc{tempo1}
  and can be emulated using the `T2' model.  For instance, the `DD'
  model is more general than the earlier Blandford \& Teukolsky
  (1976)\nocite{bt76} model (`BT'). By making various simplifying
  assumptions (see Paper~II), the `BT' model can be obtained from the
  `T2' model.  Recently it has been shown (Kramer \& Wex, private
  communication) that the `DD' model provides poor uncertainties for
  measurements of the orbital Shapiro delay when the orbital
  inclination $i \sim 90^\circ$. In the `T2' model (and the `DDS'
  model) a new parameter $x \equiv -\ln (1-\sin i)$ can be used for
  such edge-on binary systems. As shown by Kramer (private
  communication) \textsc{tempo2} provides more reliable estimates of
  the uncertainty on the new parameter $x$ (known as \textsc{shapmax}
  as it relates to the maximum Shapiro delay for a near-circular
  orbit) than on $\sin i$ for near edge-on binary systems.  For
  instance, with our observations of PSR~J1909$-$3744, the `DD' model
  gives a companion mass of $m_c = 0.2061(16)$\,M$_\odot$ and $\sin i
  = 0.99820(8)$.  The `T2' model gives $m_c = 0.2063(17)$\,M$_\odot$
  and $x = 6.28(6)$ implying that $\sin i =
  0.99813(11)$. Figure~\ref{fg:1909_m2sini} gives a $\chi^2$ plot
  indicating the fitted parameters obtained using the `T2' and `DD'
  models.

  The `T2' model, by default, returns theory-independent results.
  However, it is possible to assume that general relativity applies,
  emulating the `DDGR' model developed by Taylor (1987)\nocite{tay87}
  and Taylor \& Weisberg (1989)\nocite{tw89} for \textsc{tempo1}. This
  allows the determination of the mass of the pulsar and its
  companion.

  For wide-orbit binary systems, Kopeikin (1995)\nocite{kop95} showed
  that orbital timing parallaxes are measurable. Kopeikin
  (1996)\nocite{kop96} also described secular variations of orbital
  parameters due to the system's proper motion.  These terms are
  included as part of the `T2' model, but require an estimate of the
  longitude of the ascending node of the binary's orbit, $\Omega$.

  The PSR~B1259$-$63 system of a pulsar orbiting a Be star has proved
  difficult to model.  Wang, Johnston \& Manchester
  (2004)\nocite{wjm04} therefore developed the modified BT model that
  allowed for jumps in the Keplerian parameters at specified times
  (specifically, for this system, at periastron); such jumps are
  available in the `T2' model.  A more physical approach to the effect
  of non-point-mass companions was developed by Wex
  (1998)\nocite{wex98} through alterations in the secular behaviour of
  the longitude of periastron and the projected semi-major axis.
  These changes are also available as part of the `T2' binary model.

  For orbits with small eccentricities, the `T2' model, by default,
  produces highly covariant values for the epoch and the longitude of
  periastron.  The `ELL1' model of Wex (1998 unpublished work; see
  Lange et al. 2001\nocite{lcw+01}) was developed for pulsars in such
  orbits.  \textsc{Tempo2} provides a tool to convert between a timing
  model based on the BT model and one based on the ELL1 model which is
  defined by the parameters
  \begin{eqnarray}
    {\rm EPS1} &=& e \sin\omega \\
    {\rm EPS2} &=& e \cos\omega \\
    {\rm TASC} &=& T_0 - \frac{\omega}{2\pi} P_b 
  \end{eqnarray}
  where $e$ represents the orbital eccentricity, $\omega$ the longitude of
  periastron, $T_0$ the epoch of periastron and $P_b$ the orbital
  period. If these parameters are included in a `T2' parameter file
  then the `T2' model will emulate the `ELL1' model.


  These binary models provide the ability to determine the Keplerian
  parameters of the orbit and various post-Keplerian parameters
  (listed in Table~\ref{tb:binary}).  From the fitted values,
  \textsc{tempo2} can provide various derived quantities including the
  mass function
  \begin{equation}
   f = \frac{(m_c \sin i)^3}{(m_p + m_c)^2} =
   \frac{4\pi^2}{G}\frac{(a_p\sin i)^3}{P_b^2} {\rm .}
  \end{equation}  
  By assuming a typical pulsar mass of $m_p = 1.35 M_\odot$, a lower
  limit on the companion mass can be estimated by assuming that the
  orbit is viewed edge-on ($i = 90^\circ$), a median mass ($i =
  60^\circ$) and an upper bound at the 90\% confidence level from $i =
  26.0^\circ$ (see, e.g., Lorimer \& Kramer 2005).  These values are
  obtained by solving the mass function for the companion mass, $m_c$,
  using a Newton-Raphson method.  If $\sin i$ and $m_c$ are known
  (e.g. from Shapiro delay measurements) then the mass function gives
  the pulsar mass.

  \begin{table*}
   \caption{Binary orbital parameters included in 
   \textsc{tempo2}. }\label{tb:binary}
   \begin{tabular}{lll} \hline
   Parameter & Description & Symbol  \\ \hline
   PB        & Binary period of pulsars (d) & $P_b$  \\
   ECC       & Eccentricity of orbit    & $e$ \\
   A1        & Projected semi-major axis of orbit (lt-s) & $x$\\
   T0        & Epoch of periastron (MJD) & $T_0$ \\
   OM        & Longitude of periastron (deg)  & $\omega$ \\
   TASC      & Epoch of ascending node (MJD)  & $T_{\rm asc}$ \\
   EPS1      & $e\sin \omega$               & $\eta$ \\
   EPS2      & $e\cos \omega$               & $\kappa$ \\ 
   KOM       & longitude of the ascensing node & $\Omega$ \\
   KIN       & inclination angle               & $i$  \\
   SHAPMAX   & $-\ln(1-\sin i)$    & $s_x$ \\
   \\
   OMDOT     & Periastron advance (deg/yr)       & $\dot{\omega}$ \\
   PBDOT     & 1st time derivative of binary period & $\dot{P_b}$  \\
   ECCDOT    & Rate of change of eccentricity (s$^{-1}$) & $\dot{e}$ \\
   A1DOT     & Rate of change of semimajor axis (lt-s s$^{-1}$)& $\dot{x}$  \\
   GAMMA     & Post-Keplerian `gamma' term (s) & $\gamma$ \\
   XPBDOT    & Rate of change of orbital period minus GR prediction \\
   EPS1DOT   & Rate of change of EPS1 & $\dot{\eta}$ \\
   EPS2DOT   & Rate of change of EPS2 & $\dot{\kappa}$  \\
   MTOT      & Total system mass (M$_\odot$) & $M$  \\
   M2        & Companion mass (M$_\odot$) & $m_2$ \\
   DTHETA    & Relativistic deformation of the orbit & $d_\theta$ \\
   XOMDOT    & Rate of periastron advance minus GR prediction (deg
   yr$^{-1}$) &\\
   SINI      & Sine of inclination angle & $s$ \\
   DR        & Relativistic deformation of the orbit & $d_r$ \\
   A0        & First aberration parameter & $A$  \\
   B0        & Second aberration parameter & $B$ \\
   BP        & Tensor multi-scalar parameter & $\beta^\prime$  \\
   BPP       & Tensor multi-scalar parameter & $\beta^{\prime\prime}$ \\
   AFAC      & Aberration geometric factor & \\
   \\
   BPJEP\_k   & Epoch of a step-jump in the binary parameters &  \\
   BPJPH\_k   & Size of phase jump &  \\
   BTJPB\_k   & Size of jump in orbital period &  \\
   BTJA1\_k   & Size of jump in projected semi-major axis &\\
   BTJECC\_k  & Size of jump in orbital eccentricity & \\
   BTJOM\_k   & Size of jump in longitude of periastron & \\
   \hline\end{tabular}
  \end{table*}

 \subsection{Determining pulsar parameters}

  Timing models contain the pulsar's rotational, positional and binary
  parameters at a specific epoch.  It is often useful to determine
  such parameters at a different epoch.  \textsc{Tempo2} provides
  tools to calculate the parameters at any given epoch.  The position
  ($\alpha,\delta$) in equatorial coordinates are updated from the
  position epoch using proper motion determinations ($\mu_\alpha \cos
  \delta,\mu_\delta$)
  \begin{eqnarray}
   \alpha^\prime = \alpha + \mu_\alpha \Delta t_A/\cos \delta\\\
   \delta^\prime = \delta + \mu_\delta \Delta t_A 
  \end{eqnarray}
  where $\Delta t_A$ represents the difference between the requested epoch
  and the current model position epoch.

  The pulsar's spin frequency and its first derivative are updated
  from the first and subsequent frequency derivatives

  \begin{eqnarray}
   \nu^\prime = \nu + \dot{\nu} \Delta t_P +
   \frac{1}{2}\ddot{\nu}(\Delta t_P)^2 + \ldots \\
   \dot{\nu}^\prime = \dot{\nu} + \ddot{\nu} \Delta t_P + \ldots
  \end{eqnarray}
  where $\Delta t_P$ represents the change between the requested epoch
  and the current epoch for the given frequency determinations.

\begin{table*}
 \begin{center}
 \caption{Example parameters in a \LaTeX~table format for
 PSR~J0437$-$4715 obtained using the \textsc{publish}
 plug-in.}\label{tb:latex}
\begin{tabular}{ll}
\hline\hline
\multicolumn{2}{c}{Fit and data-set} \\
\hline
Pulsar name\dotfill & J0437$-$4715 \\ 
MJD range\dotfill & 53041.3---53767.3 \\ 
Number of TOAs\dotfill & 603 \\
Rms timing residual ($\mu s$)\dotfill & 1.6 \\
 Weighted fit\dotfill &  N \\ 
\hline
\multicolumn{2}{c}{Measured Quantities} \\ 
\hline
Right ascension\dotfill &  04:37:15.78858(13) \\ 
Declination\dotfill & $-$47:15:08.4685(15) \\ 
Pulse frequency (s$^{-1}$)\dotfill & 173.68794630602(7) \\ 
First derivative of pulse frequency (s$^{-2}$)\dotfill & $-$1.7292(4)$\times 10^{-15}$ \\ 
Dispersion measure (cm$^{-3}$pc)\dotfill & 2.64123(17) \\ 
Proper motion in right ascension (mas\,yr$^{-1}$)\dotfill & 120.9(3) \\ 
Proper motion in declination (mas\,yr$^{-1}$)\dotfill & $-$71.0(3) \\ 
Parallax (mas)\dotfill & 8(3) \\ 
Orbital period (d)\dotfill & 5.7410464584(16) \\ 
Epoch of periastron (MJD)\dotfill & 51194.620(6) \\ 
Projected semi-major axis of orbit (lt-s)\dotfill & 3.36670624(19) \\ 
Longitude of periastron (deg)\dotfill & 0.9(4) \\ 
Orbital eccentricity\dotfill & 0.00001899(11) \\ 
\hline
\multicolumn{2}{c}{Set Quantities} \\ 
\hline
Epoch of frequency determination (MJD)\dotfill & 51194 \\ 
Epoch of position determination (MJD)\dotfill & 51194 \\ 
Epoch of dispersion measure determination (MJD)\dotfill & 51194 \\ 
Sine of inclination angle\dotfill & 0.6788 \\ 
First derivative of orbital period\dotfill & 3.64$\times 10^{-12}$ \\ 
Periastron advance (deg/yr)\dotfill & 0.016 \\ 
Companion mass ($M_\odot$)\dotfill & 0.236 \\ 
\hline
\multicolumn{2}{c}{Derived Quantities} \\
\hline
$\log_{10}$(Characteristic age, yr) \dotfill & 9.20 \\
$\log_{10}$(Surface magnetic field strength, G) \dotfill & 8.76 \\
\hline
\multicolumn{2}{c}{Assumptions} \\
\hline
Clock correction procedure\dotfill & TT(TAI) \\
Solar system ephemeris model\dotfill & DE405 \\
Binary model\dotfill & DD \\
Model version number\dotfill & 5.00 \\ 
\hline
\end{tabular}

Note: Figures in parentheses are twice the nominal 1$\sigma$
\textsc{tempo2} uncertainties in the least-significant digits quoted.
 \end{center}
\end{table*}

  Binary parameters are more problematic.  The epoch of periastron
  $T_0$ (or the epoch of the ascending node, $T_{asc}$) is updated to
  the closest periastron to the requested epoch by calculating the
  nearest integer $n$ to

  \begin{eqnarray}
   a_n = \frac{\Delta t_b}{P_b} - \frac{1}{2}\dot{P_b}\left(\frac{\Delta
   t_b}{P_b}\right)^2
  \label{eqn:n}
  \end{eqnarray}
  which represents the number of orbits since the timing model epoch
  of periastron $T_0$, i.e., in interval $\Delta t_b$.  The epoch of
  periastron is subsequently updated
  
  \begin{eqnarray}
   T_0^\prime = T_0 + (\Delta t_b)^\prime
  \end{eqnarray}
  where $(\Delta t_b)^\prime$ is obtained by solving
  Equation~\ref{eqn:n} with $a_n = n$.  The orbital period ($P_b$),
  longitude of periastron ($\omega$), eccentricity ($e$) and projected
  semi major axis ($a_1$) are updated as
  \begin{eqnarray}
   P_b^\prime = P_b + \dot{P_b}(\Delta t_b)^\prime \\
   \omega^\prime = \omega + \dot{\omega}(\Delta t_b)^\prime \\
   e^\prime = e + \dot{e}(\Delta t_b)^\prime \\
   a_1^\prime = a_1 + \dot{a_1}(\Delta t_b)^\prime
  \end{eqnarray}

 \subsection{Publishing timing ephemerides}

 \textsc{Tempo1} provided only limited output formats for the timing
 model parameters.  The \textsc{publish} plug-in for \textsc{tempo2}
 provides the user with the ability to produce output in a \LaTeX
 ~table format (as in Table~\ref{tb:latex} where, in this case, the
 uncertainty on each parameter for PSR~J0437$-$4715 has been
 multiplied by a factor of two and the error corresponds to the
 uncertainty on the last quoted digit).

 If many parameters are included in the fit, then it is possible that
 the resulting uncertainties are covariant.  This can be checked using
 the \textsc{matrix} plug-in that displays the correlation matrix for
 the fit (a typical output is shown in Table~\ref{tb:matrix} for
 PSR~J0437$-$4715).  It is common for low-eccentricity binary pulsars
 that the epoch of periastron and the longitude of periastron are
 near-degenerate.  It is common in publications to quote highly
 covariant parameters with more precision than suggested by the formal
 errors (see for instance Ryba \& Taylor 1991)\nocite{rt91b}, however,
 the use of the ELL1 binary model is preferred.

 Published timing parameters are used to make predictions for on-line
 observations of pulsars, for comparing results between different
 observing systems and for searching for variations in the parameters
 with time.  It is therefore essential that full details of how the
 timing model was created are published alongside the parameters.  For
 any published timing parameters it is necessary to 1) indicate
 whether the uncertainties represent 1 or 2$\sigma$ formal errors on
 the fitted parameters and whether a weighted or non-weighted fitting
 procedure was used, 2) specify the Solar System ephemeris used, 3)
 indicate the \textsc{tempo2} version number and describe which of the
 default \textsc{tempo2} algorithms have not been used and which
 non-default algorithms included, 4) provide full information of any
 pre-whitening carried out on the dataset, 5) define the coordinate
 system used and 6) provide details of the clock correction process.
 The fitting process should be iterated until the pre- and post-fit
 values are identical indicating that the fit has converged. If a
 weighted fit is carried out then it is necessary, in order to obtain
 accurate errors on the fitted parameters, to ensure that the reduced
 $\chi^2$-value of the fit is close to unity. It is also common
 practice and desirable to convert the measured parameters so that the
 period, position and binary epochs refer to the centre of the data
 span.

\section{Analysing the timing residuals}

 Any systematic feature in the post-fit timing residuals indicates
 that some effect is not being described by the timing
 model. Analysing such features potentially provides important
 information about various perturbations, including the presence of
 planetary companions, variations in the interstellar medium,
 precession of the neutron star or irregularities in the pulsar's
 rotation and spin-down.  The timing residuals can also provide an
 indication that the TOAs are affected by calibration problems or
 other instrumental effects.

 The pre- and post-fit timing residuals can be plotted in numerous
 ways.  For instance, the \textsc{plk} plug-in allows the user to plot
 the residuals versus parameters such as day, observing frequency,
 binary phase, observation length or the parallactic angle.  Other
 plug-ins allow the user to obtain a list of the timing residuals,
 barycentric arrival times, clock corrections etc. in a specified
 tabular format.

 As the true uncertainty on any fitted parameter is the combination of
 the random and systematic errors, it is necessary to attempt to
 quantify the effects of the systematic errors present in the data.
 Ryba \& Taylor (1991)\nocite{rt91b} discuss two methods to do this
 which are implemented in the \textsc{errors} plug-in

\begin{figure}
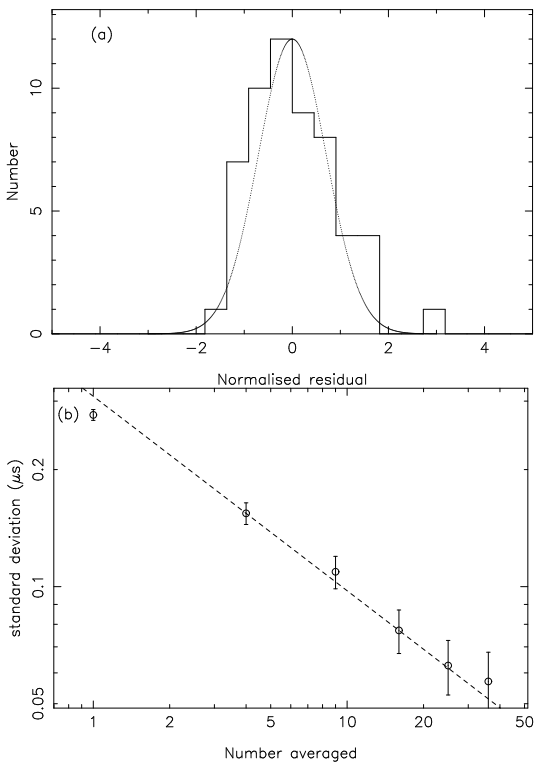

   \begin{center}
    \includegraphics[angle=-90,width=7cm]{res1.ps}
    \includegraphics[angle=-90,width=7cm]{res2.ps}
   \end{center}
  \caption{a) Normalised residuals for PSR~J1909$-$3714 plotted as a
  histogram with a Gaussian curve overlaid. b) Rms timing residuals
  after averaging different numbers of residuals together with a line
  indicating the expected $\sqrt N$ dependence.  These plots were
  created using the \textsc{errors} plug-in.}\label{fg:residuals}
 \end{figure}

 \begin{enumerate}
  \item{it is possible to plot a histogram of the normalized postfit
    residuals (the value of the residual divided by its estimated
    uncertainty).  If this histogram follows a Gaussian distribution
    then it is likely that the data are not significantly affected by
    systematic effects (Figure~\ref{fg:residuals}a).}
   \item{compute averages of consecutive sets 1, 2, 4, 8 and 16
    normalised residuals and plot the standard deviations within each
    group.  Random Gaussian measurement errors produce deviations
    $\propto 1/\sqrt{N}$ where $N$ is the number of residuals
    averaged (Figure~\ref{fg:residuals}b).}
 \end{enumerate}

 \textsc{Tempo2} plug-ins provide numerous tools for analysing the
 post-fit timing residuals or for outputting the residuals in formats
 that are suitable for other data-analysis packages.  We have noted
 that large numbers of packages have been created to analyse timing
 residuals from the output files of the \textsc{tempo1} software and
 therefore provide facilities within \textsc{tempo2} to produce output
 files with the same format as \textsc{tempo1}. However, it is
 relatively straightforward to develop new plug-ins or to convert old
 software for use by \textsc{tempo2}. For instance, a template plug-in
 is available which users can modify for their own specific uses. Many
 such plug-ins to analyse the post-fit timing residuals are currently
 being designed or tested.  Examples include the implementation of a
 multi-resolution CLEAN deconvolution algorithm and periodicity
 searches (M. Rissi; private communication).

 \begin{table*}
  \begin{center}
  \caption{The correlation matrix for the fitted parameters for
  PSR~J0437$-$4715 data obtained using the \textsc{matrix} plug-in for
  \textsc{tempo2}.  The global correlation (gcor) parameter is a
  measure for the strongest correlation between the fitted variable
  and a linear combination of all other variables.  ${\rm dp} =
  -\log_{10}(1-{\rm gcor}^2)^{1/2}$ provides an estimate of the number
  of ``insignificant'' digits that should be quoted in a timing
  solution; see text.}\label{tb:matrix}
  \begin{verbatim}
          F0              T0              A1              OM              ECC      
F0      +1.00000000
T0      -0.13474657     +1.00000000
A1      +0.11393821     +0.00841676     +1.00000000
OM      -0.13474630     +0.99999998     +0.00841640     +1.00000000
ECC     +0.00709055     +0.18355036     +0.10675629     +0.18355054     +1.00000000
------------------------------------------------------------------------------------
gcor    +0.17835063     +0.99999998     +0.15566451     +0.99999998     +0.21251747
dp           0.1             7.8             0.1             7.8             0.1
------------------------------------------------------------------------------------
   \end{verbatim}  
 \end{center}
 \end{table*}

\section{Predictive mode}

  The folding of pulsar signals proceeds on the basis of the predicted
  time evolution of the phase of the pulse train incident upon the
  observatory. The timing model specifies this evolution, but is too
  computationally intensive for real-time applications.  Like
  \textsc{tempo1}, \textsc{tempo2} is able to produce a polynomial
  approximation of the phase, $\phi(t)$, and pulse frequency, $\nu$,
  over specified time intervals.  The number of coefficients and the
  time span fitted can be set by the user; \textsc{tempo2} provides a
  warning message if the rms deviation between the model and the data
  is large and more coefficients or a shorter span are necessary.


\subsection{\textsc{Tempo1}-compatible polynomials}

To ease the transition from \textsc{tempo1} to \textsc{tempo2}, a
facility is provided to produce predictive polynomials in the same
format as those made by \textsc{tempo1}. These consist of a series of
sets of coefficients of polynomials that approximate the evolution of
pulse phase incident upon the specified observatory.  Owing to
interplanetary and interstellar dispersion, the polynomial is specific
to a given observing frequency. For observations at radio wavelengths,
there is usually a significant variation of instantaneous pulse phase
across the observing band. If high precision is required, the form of
this variation is dependent on many parts of the full timing model,
thereby defeating any gains in simplicity offered by substituting a
polynomial form for the time evolution. For this reason,
\textsc{tempo2} can produce new time- and frequency-dependent
predictive polynomials (Section \ref{sec:2dpoly}) which are
recommended for precision applications.

A simplified approximation to the frequency dependence may
be used in less critical applications. For isolated pulsars, the
difference in time of emission between radiation received simultaneously at
an observing frequency, $f$, and the frequency for which the polynomial
applies, $f_0$, is:
\begin{equation}
\Delta_D = D\left(f^{-2}-f_0^{-2}\right),
\end{equation}
where $D$ is the dispersion constant (Section
\ref{sec:dm}). Neglecting any variation in pulse frequency over the
interval $\Delta_D$ (safe for isolated pulsars), the resultant phase
difference is simply $\nu\Delta_D$, where $\nu$ is the pulse frequency
at the epoch of consideration. The pulse frequency in the reference
frame of the observatory is simply obtained from the polynomial, and
the observing frequency is also generally known by its value in the
observatory frame. It is therefore convenient to perform the
calculation in this frame, after transforming the dispersion constant
($D$, with dimensions of time$^{-1}$) using the ``Doppler
shift''\footnote{In \textsc{tempo2} the rate correction also includes
gravitational redshift and time dilation.} value ($\beta$) specified
in the polynomial data file:
\begin{equation}
\phi(t,f) \simeq \phi(t,f_0) - \nu(t, f_0)
\left(f^{-2}-f_0^{-2}\right)\frac{D}{1+\beta}.
\label{eq:dedisp}
\end{equation}

 \noindent To the best of our knowledge, the application of $\beta$ to
the dispersion constant is often neglected in existing dedispersion
software, incurring an error of up to $\sim\Delta_D \cdot 10^{-4}$,
typically of the order of several microseconds.  It should also be
noted that unlike \textsc{tempo2}, the value of the dispersion measure
provided in \textsc{tempo1} polynomial files does not include the
interplanetary contribution, typically in the range 100~ns --
100~$\mu$s.

\begin{figure}
  \begin{center}
    \includegraphics[width=7cm]{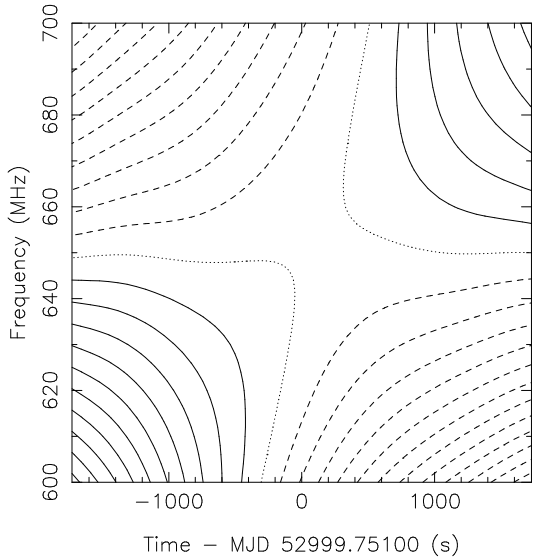}
  \end{center}
  \caption{Contour map of the difference between the pulse phase as
predicted by the full \textsc{tempo2} timing model, versus a
single-frequency polynomial approximation in conjunction with Eq.\
\ref{eq:dedisp}, for an hour-long observation of the binary pulsar PSR
J1906+0746 with $f_0=650$~MHz. Contours are spaced by 5~ns; dashed
contours are negative, positive contours are solid and the zero contour
is dotted. Two main effects are noticeable. The largest effect is due to the
evolution in the observatory Doppler shift, manifested as a monotonic
function of time and frequency (delay $\propto t(f^{-2} - f_0^{-2})$,
where $t$ is the time relative to the observation midpoint). The
second, smaller effect is due to second- and higher-order terms in the
orbital motion of the pulsar (delay approximately $\propto (f^{-2}
-f_0^{-2})^2$). At the midpoint in time where the first effect vanishes,
the second effect appears as a form roughly quadratic in frequency offset.
Elsewhere it contributes to the overall assymmetry of the difference function.}
\label{fig:1906_polyco_errors}
 \end{figure}

For isolated pulsars, the limiting factor in the accuracy of this
approach is the time evolution of the transformation between the
observatory and barycentric frames ($1+\beta$).  Only a single value
is provided for each polynomial, which typically spans several
hours. The rotation of the Earth accelerates the observatory
sufficiently to change $\beta$ at rates of up to $\sim
10^{-10}$~s$^{-1}$, resulting in errors of the order of tens of
nanoseconds. Figure~\ref{fig:1906_polyco_errors} illustrates the
effect. 

\begin{figure}
  \begin{center}
    \includegraphics[width=7cm]{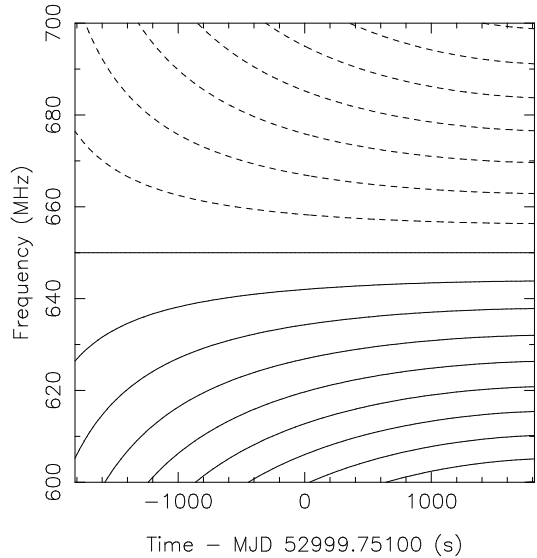}
  \end{center}
  \caption{Contour map of the fractional difference in apparent pulse
frequency at a given observing frequency, versus the central frequency
(650 MHz), for an hour-long observation of the binary pulsar PSR
J1906+0746. Contours are spaced at $10^{-8}$, corresponding to 10 ns
of smearing per second of integration. Dashed contours are negative. }
\label{fig:1906_polyco_delta_f}
 \end{figure}

For binary pulsars, more severe effects are encountered, owing to the
different orbital phases of emission of radiation received
simultaneously at different frequencies. The most striking result of
this is that, for observing frequencies well away from the frequency
assumed in the construction of the polynomial, pulse profile peaks
lead or lag the prediction as a function of orbital phase. While
eye-catching, to the extent that the error in the predicted pulse
phase is constant over the course of a given integration, this effect
should not manifest directly in the measured pulse arrival
times. Likewise, if pulse profile data are
combined from several frequencies (or an observing band is coherently
dedispersed), as long as ``dedispersion'' proceeds as prescribed in
Eq.\ \ref{eq:dedisp}, the apparent orbital-phase-dependent effect is
compensated to first order by the variation in the topocentric pulse
frequency used to convert a dispersion time delay to a phase
delay. However, in addition to the secular drift of $\beta$, two
remaining effects dictate the use of time- and frequency-dependent
two-dimensional polynomials for precision timing of binary pulsars.

\begin{table*}
 \centering
 \begin{minipage}{140mm}
  \caption{Smearing and phase errors with \textsc{tempo1}-style predictive polynomials for sample pulsars}
  \label{tab:smear}
  \begin{tabular}{@{}lrrrrrrrrrr@{}}
  \hline Name & $P$  & DM  & $a \sin i$  &
   $P_{\rm orb}$  %
& $\Delta_1$\footnote{Maximum smearing due to pulse frequency offset at $f=600$~MHz, vs $f_0=650$~MHz, over 100-second integration (see text). Note, values are approximate only: orbital eccentricity neglected.} %
& $\Delta_1/P$  %
& $\Delta_2$\footnote{Maximum timing error due to neglected order-$\Delta_D^2$ term, for $f=600$~MHz, $f_0=650$~MHz (see text)} %
& $\Delta_2/P$ \\
 & (ms) & (cm pc$^{-3}$) & (lt-s) & (d) & (ns) & ($10^{-6}$) & (ns) &($10^{-6}$) \\
 \hline
   J1906+0746 &  144.1 & 217.8 &  1.42 &  0.17 &  10100 & 70 & 18.8 &0.13\\ 
J0737$-$3039A &   22.7 &  48.9 &  1.42 &  0.10 &  6000 & 260 & 2.5 &0.11\\ 
  B1744$-$24A &   11.6 & 242.2 &  0.12 &  0.08 &  4600 & 400 & 9.4 & 0.8\\ 
     B1913+16 &   59.0 & 168.8 &  2.34 &  0.32 &  3400 & 58 & 4.9 & 0.08\\ 
 J1756$-$2251 &   28.5 & 121.2 &  2.76 &  0.32 &  2900 & 100 &  3.0 & 0.11\\ 
 J1802$-$2124 &   12.6 & 149.6 &  3.72 &  0.70 &  1000 & 81 & 1.3 & 0.10\\ 
 J1435$-$6100 &    9.3 & 113.7 &  6.18 &  1.35 &    345 &  37 & 0.3 & 0.04 \\ 
   J0218+4232 &    2.3 &  61.3 &  1.98 &  2.03 &     26 &  11 & $14\times 10^{-3}$ & $6\times 10^{-3}$\\
     B1957+20 &    1.6 &  29.1 &  0.09 &  0.38 &     16 &  10 & $4 \times 10^{-3}$ & $ 2\times 10^{-3}$\\
 J1909$-$3744 &    2.9 &  10.4 &  1.90 &  1.53 &      7.6 & 2.6 & $6 \times 10^{-4}$ & $2 \times 10^{-4}$\\
 J2145$-$0750 &   16.1 &   9.0 & 10.16 &  6.84 &      1.8 & 0.11 & $14 \times 10^{-5}$ & $8 \times 10^{-6}$\\
    B1855+09 &    5.4 &  13.3 &  9.23 & 12.33 &      0.7 & 0.14 & $8 \times 10^{-5}$ & $15 \times 10^{-6}$\\
 J0437$-$4715 &    5.8 &   2.6 &  3.37 &  5.74 &     0.2 & 0.04 & $1 \times 10^{-5}$ & $10^{-6}$ \\
\hline
\end{tabular}
\end{minipage}
\end{table*}

Firstly, unless taken into account, the variation in apparent pulse
frequency as a function of observing frequency can be large enough to
cause appreciable smearing of the pulse profile if not taken into
account. With scintillation, the smearing may be significantly
asymmetric, resulting in errors in measured pulse times of
arrival. The fractional change in pulse frequency is given to first
order by $\Delta_D a/c$ where $a$ is the line-of-sight orbital
acceleration. See Figure \ref{fig:1906_polyco_delta_f} for an example.
Smearing widths for sample pulsars and observing configurations are
listed in Table \ref{tab:smear}.

Secondly, if profiles from several frequencies are to be combined (or
an observing band is to be coherently dedispersed), Eq.\
\ref{eq:dedisp} corrects only to first order in $\Delta_D$. Since the
topocentric frequency, $\nu(t, f)$ itself varies with dispersion
delay, the second order term (expressed as a time delay) is given by
$\Delta^2_D a/2c$. This effect is apparent in Figure
\ref{fig:1906_polyco_errors}, especially around $t=0$ where the
differential Doppler effect vanishes. The magnitudes of the effect for
sample pulsars and observing configurations are provided in Table
\ref{tab:smear}.

\subsection{Time- and Frequency-dependent predictions}
\label{sec:2dpoly}
To overcome the limitations of the \textsc{tempo1}-type
predictive polynomial described in the previous section,
\textsc{tempo2} is able to compute a two-dimensional
polynomial approximation to the timing model, with time and
observing frequency as its arguments. This mode is recommended
for precision applications in future processing systems.

The polynomial is expressed
in terms of a two-dimensional adaptation of the conventional Chebyshev
basis functions:
\begin{equation}
p_{ij}(x, y) = \cos(i \cos^{-1} x) \cos(j \cos^{-1} y).
\end{equation}
\noindent Analogous to the one-dimensional case, a discrete form of
a set of such basis functions is orthogonal if computed on a grid
of $M\times N$ coordinates $(x_i, y_j)$ satisfying
\begin{equation}
\cos(M \cos^{-1} x_i)\cos(N \cos^{-1} y_j)= 0.
\label{eq:cheby_ortho_condition}
\end{equation}
The coefficients of the polynomial approximation to a function $g(x,y)$
are therefore easily computed via the inner product:
\begin{equation}
c_{kl} = \frac{4}{MN}\sum_{i=1}^{M} \sum_{j=1}^{N} p_{kl}(x_i, y_j) g(x_i, y_j).
\end{equation}
This yields an approximation function
\begin{equation}
g(x, y) \simeq \sum_{k=1}^{M} \sum_{l=1}^{N} c_{kl} p_{kl}(x,y),
\end{equation}
which is exact for $x, y$ coordinates satisfying 
Eq.\ \ref{eq:cheby_ortho_condition}. For computation, this can
be rewritten as a one-dimensional Chebyshev polynomial with
another Chebyshev polynomial defining its coefficients:
\begin{equation}
g(x, y) \simeq \sum_{l=1}^{N} T_l(y) \sum_{l=1}^{N} T_k(x) c_{kl},
\end{equation}
\noindent where the one-dimensional basis functions, $T_k(x) = \cos(k
\cos^{-1} x)$, can be computed efficiently using Clenshaw's 
recurrence relation \cite{ptvf92}. 

After mapping the requested intervals in time and observing frequency
to $x$ and $y$ in the interval $[-1,1]$, \textsc{tempo2} computes the
coefficients $c_{kl}$ approximating the function $\phi(t, f)-kf^{-2}$,
where $k$ is a constant computed to remove the bulk of the frequency
dependence due to interstellar dispersion.  Subtraction of this term
significantly reduces the number of coefficients needed for an accurate
approximation. The $M\times N$ coefficients and the constant $k$ are
written to file to allow the later construction of the approximation
of $\phi(t, f)$ without reference to the full timing model.

Using this approach, the timing model, including its frequency
dependence, can be approximated to sub-nanosecond accuracy using a
modest number of coefficients. As an example, the phase evolution of
PSR J1906+0746 over the time and frequency interval depicted in
Figure\ \ref{fig:1906_polyco_errors} requires nine coefficients in the
time axis and six in frequency, yielding a difference function with an
rms variation of 450\,ps. For an observation of an isolated pulsar,
over the same time interval and frequency band, only $5\times3$
coefficients are needed to model the time- and frequency-dependent
variations in phase, which apart from the basic pulsar frequency, are
dominated by the acceleration of the observatory due to the rotation
of the earth.

The software library packaged with \textsc{tempo2} provides routines
for evaluating the predicted pulse phase and pulse frequency as a
function of time and observing frequency. For convenience in
applications involving folding of pulsar time-series data, the
software can also produce a piecewise linear approximation which
minimises the mean error and keeps the rms within specified bounds. A
plugin called \textsc{polytest} is also provided to assess the accuracy
of a computed polynomial (either 1- or 2-dimensional), reporting on
the rms and extrema of the residuals and producing plots such as those
in Figures \ref{fig:1906_polyco_errors} and
\ref{fig:1906_polyco_delta_f}.

 \section{Summary and conclusion}

 In summary we list below the improved features of \textsc{tempo2}
 compared to \textsc{tempo1}.  For coordinate systems, propagation
 delays and ephemerides, \textsc{tempo2}

  \begin{itemize}
   \item{is compliant with IAU 2000 resolutions and implements updated
         precession and nutation models, polar motion, the ICRS
         coordinate system and TCB (SI units) instead of TDB;}
   \item{corrects the observing frequency for relativistic time
         dilation;}
   \item{uses an improved tabulation of the Solar System Einstein delay;}
   \item{includes atmospheric propagation delays;}
   \item{includes the Shapiro delay due to the planets; and}
   \item{includes the second-order Shapiro delay due to the Sun.}
  \end{itemize}

 For the fitting routines, \textsc{tempo2}

  \begin{itemize}
   \item{has the ability to simultaneously fit to the timing
        residuals of multiple pulsars;}
   \item{implements frequency-dependent fitting (not only DM delays);}
   \item{has the ability to fit for DM at each epoch using simultaneous observations;}
   \item{allows fits for an arbitrary number of pulse frequency
   derivatives, dispersion measure derivatives and glitches;}
   \item{simplifies and provides flexible methods for placing
         arbitrary offsets between TOAs obtained at different
         frequencies or observatories;}
   \item{can whiten data using harmonically related sinusoids;}
   \item{provides a brute-force method for obtaining timing solutions;}
   \item{fully includes the effects of secular motion of the pulsar; and}
   \item{includes the orbital parallax terms in binary models.}
  \end{itemize}

 The predictive mode in \textsc{tempo2} provides both time and
 frequency-dependent predictions and therefore deals correctly with
 \begin{itemize}
  \item{Earth rotation and}
  \item{binary motion.}
 \end{itemize}

 Other miscellaneous improvements include
 
  \begin{itemize}
   \item{generalised input formats;}
   \item{generalised output formats;}
   \item{the ability to simulate pulse arrival times;}
   \item{the implementation of numerous graphical interfaces;}
   \item{calculating the post-fit residuals (instead of predicting them);}
   \item{the ability to update binary parameters to a given epoch and}
   \item{generalised clock correction routines.}
  \end{itemize}

  We have presented a new software package, \textsc{tempo2}, that
  supersedes the existing \textsc{tempo} package.  \textsc{Tempo2} has
  been analysed in detail and we believe that all the corrections to
  the measured TOAs that have been implemented are accurate to better
  than 1\,ns.  The software has been designed so that it is easy to
  modify current routines and to add new functions to describe
  phenomena which affect pulsar timing residuals.  For instance, a
  user can easily implement a new binary model or create a personal
  graphical interface.

  The \textsc{tempo2} framework is such that it is relatively easy to
  fit global parameters across multiple data sets.  This will be the
  topic of a forthcoming paper and will be used in the hunt for
  gravitational waves, refining the Solar System ephemeris and
  establishing a pulsar-based timescale.

\section*{Acknowledgements}

The \textsc{tempo2} package is based on the original \textsc{tempo}
Fortran code.  This software was developed over many years by multiple
authors including J. Taylor, R. Manchester, D. Nice, W. Peters,
J. Weisberg, A. Irwin and N. Wex.  We thank the Australian pulsar
community, J. Weisberg, D. Lorimer and M. Kramer for suggestions and
comments on \textsc{tempo2}.  The data presented in this paper were
obtained as part of the Parkes Pulsar Timing Array project that is a
collaboration between the ATNF, Swinburne University and The
University of Texas, Brownsville, and we thank our collaborators on
this project.  The Parkes radio telescope is part of the Australia
Telescope which is funded by the Commonwealth of Australia for
operation as a National Facility managed by CSIRO.

\bibliography{tempo2,modrefs,psrrefs,crossrefs}
\bibliographystyle{mn}

\appendix

\section{List of plug-in packages available for \textsc{tempo2}}

\begin{itemize}

 \item{\textsc{Basic}, plots a ${\rm P}-\dot{{\rm P}}$ diagram and indicates
 the position of the pulsar being analysed. Options are available to
 also display a sky-projection that indicates the pulsar's position
 and derived parameters such as the pulsar's characteristic age and
 surface dipole magnetic field strength are determined.}

 \item{\textsc{Compare}, accepts two input parameter files and provides
 routines to compare differences between the residuals obtained using
 the two models. For instance, this interface can be used to compare
 the effects of different clock or ephemeris files or
 different binary models.}

 \item{\textsc{CompareRes}, accepts two input arrival-time files and
 provides routines to compare differences between the residuals.}

 \item{\textsc{Delays}, shows the \textsc{tempo2} calculated delays being added
 to the measured arrival times. For instance, clock corrections,
 ephemeris delays and dispersion delays are included.}

 \item{\textsc{Errors}, used to study systematic and random errors in
 the timing residuals.}

 \item{\textsc{Fake}, allows the user to create simulated arrival times that
 fit a given timing model.  The addition of red and white-noise is
 possible.}

 \item{\textsc{General}, a user-specified output format for
 the pre- and post-fit parameters.}

 \item{\textsc{General2}, a user-specified output format for
 displaying the site and barycentric arrival times, the timing
 residuals and various clock and propagation corrections.}

 \item{\textsc{Gorilla}, finds timing solutions using a brute-force
 method.}

 \item{\textsc{List}, provides a listing of the arrival times,
 residuals, clock corrections and propagation delays.}

 \item{\textsc{Matrix}, displays the correlation matrix of the fitted parameters.}

 \item{\textsc{Plk}, an interface that plots the timing residuals versus
 parameters such as day, binary phase, length of observation
 etc. Various functions are available including the ability to
 highlight selected points, view pulse profiles and to delete
 observations and refit the data.}

 \item{\textsc{Polytest}, provides diagnostics of the approximation error
 of a predictive polynomial, including minimum, maximum and rms and 
 frequency- and time-dependent contour maps.}

 \item{\textsc{Publish}, produces publication-quality \LaTeX ~tables of the
 parameters.}

 \item{\textsc{Spectral}, provides basic spectral analysis tools for the
 timing residuals - including periodograms, auto-correlation functions
 and CLEAN deconvolution.}

 \item{\textsc{Splk}, allows the plotting of multiple pulsar timing
 residuals simultaneously.}

 \item{\textsc{Stats}, output mode that provides basic information
 about the pulsar and its fit.  This plug-in gives the rms residuals
 for each observing frequency available.}

 \item{\textsc{Stridefit}, allows fitting of subsets of adjacent
 observations. The resulting fit parameters can be stored and
 subsequently plotted versus time.}

 \item{\textsc{Transform}, transforms a timing model created using
 \textsc{tempo1} to one using the TCB timescale and hence suitable for
 use with \textsc{tempo2}.}

\end{itemize}

\end{document}